\documentclass[aps,prx,twocolumn,groupedaddress,sort&compress,numbers]{revtex4-2}
\usepackage{graphicx}

\usepackage{subcaption}
\usepackage{dcolumn}
\usepackage{bm}
\usepackage[utf8]{inputenc}
\usepackage{array, float, tabularx, lipsum, amsmath, multirow} 
\usepackage{siunitx, xcolor}
\usepackage[version=4]{mhchem}
\graphicspath{{figs/}{figsgaoerb/}} 
\usepackage[colorlinks,linkcolor=blue,anchorcolor=blue,citecolor=blue]{hyperref}
\usepackage{svg}
\usepackage{ragged2e}
\usepackage{caption} 
\usepackage{booktabs}

\usepackage{orcidlink}



\begin{document}

\title{Continuous-mode analysis for practical continuous-variable quantum key distribution}



\author{Yanhao Sun\orcidlink{0009-0008-1579-0216}$^{1}$}
\author{Jiayu Ma\orcidlink{0009-0006-0886-7751}$^{1}$}
\author{Xiangyu Wang\orcidlink{0000-0003-2132-0065}$^{1}$}
\email{xywang@bupt.edu.cn}
\author{Song Yu$^{1}$}
\author{Ziyang Chen\orcidlink{0000-0002-8343-8029}$^{2}$}
\email{chenziyang@pku.edu.cn}
\author{Hong Guo\orcidlink{0000-0003-0644-6698}$^{2}$}

\affiliation{$^{1}$State Key Laboratory of Information Photonics and Optical Communications, Beijing University of Posts and Telecommunications, Beijing 100876, China.}
\affiliation{$^{2}$State Key Laboratory of Photonics and Communications, School of Electronics, and Center for Quantum Information Technology, Peking University, Beijing 100871, China.}


\date{\today}

\begin{abstract}
Continuous-variable quantum key distribution (CV-QKD) enables two remote parties 
to establish information-theoretically secure keys and offers high practical feasibility 
due to its compatibility with mature coherent optical communication technologies. 
However, as CV-QKD systems progress toward digital implementations, 
device nonidealities drive the optical field from a single-mode to a continuous-mode region, 
thereby underscoring the mismatch between theoretical models and practical systems. 
Here, we introduce temporal modes to construct an entanglement-based scheme 
that more accurately captures device nonidealities 
and develop a corresponding secret key rate calculation method 
applicable to continuous-mode scenarios. 
We demonstrate that optimizing the pulse-shaping format 
can significantly improve performance under detector-bandwidth-limited conditions. 
Experimental results also confirm that 
the proposed model effectively describes the impact of sampling-time deviations. 
We further analyze a 
linear weighted-reconstruction digital signal processing method, 
which improves the secret key rate by approximately 50\% in a 30-km fiber experiment 
without requiring additional hardware, demonstrating a substantial performance enhancement 
at metropolitan distances. The proposed theoretical framework accommodates 
a broader range of experimental conditions and can guide the optimization of 
digital CV-QKD systems.
\end{abstract}

\maketitle


\section{\label{I. Introduction}Introduction}
Quantum key distribution (QKD)~\cite{R1_QKD1,R2_QKD2,R3_QKD3,R4_QKD4,R5_QKD5} 
relies on the principles of quantum mechanics to provide 
information-theoretically~\cite{R6_InformationSecurity} secure keys, 
offering an effective defense against the risks introduced by 
quantum computing~\cite{R7_Shor} to classical cryptographic systems.
Among various QKD implementations, continuous-variable (CV) QKD~\cite{R8_CVQKD1,R9_CVQKD2} has attracted 
significant interest because it can be built using commercially available 
optical communication components. 
In recent years, substantial progress has been made in the theory
~\cite{R10_CVQKD_T1,R11_CVQKD_T2,R12_CVQKD_T3,R13_CVQKD_T4,R14_CVQKD_T5,R15_CVQKD_T6,R16_CVQKD_T7,R17_CVQKD_T8,R18_CVQKD_T9,R19_CVQKD_T10,R20_CVQKD_T11,R21_CVQKD_T12,R22_CVQKD_T13}, 
experimental demonstrations 
~\cite{R23_CVQKD_E1,R24_CVQKD_E2,R25_CVQKD_E3,R26_CVQKD_E4,R27_CVQKD_E5,R28_CVQKD_E6,R29_CVQKD_E7,R30_CVQKD_E8,R31_CVQKD_E9,R32_CVQKD_E10,R33_CVQKD_E11,R34_CVQKD_E12,R35_CVQKD_E13,R36_CVQKD_E14}, 
post-processing techniques
~\cite{R37_CVQKD_P1,R38_CVQKD_P2,R39_CVQKD_P3,R40_CVQKD_P4,R41_CVQKD_P5} , 
and network deployment 
~\cite{R42_CVQKD_N1,R43_CVQKD_N2,R44_CVQKD_N3,R45_CVQKD_N4}
of CV-QKD. 
To further leverage its compatibility with existing fiber-optic communication systems, 
researchers have introduced advanced digital signal processing (DSP) 
~\cite{R46_DSPstage1,R47_DSPstage2,R48_DSPstage3}
techniques from 
classical coherent communication, pushing CV-QKD steadily progress
toward digital implementations.

However, in practical CV-QKD systems, 
the communicating parties (typically referred to as Alice and Bob) 
are often constrained by device nonidealities, 
causing the optical field to evolve from an idealized single-mode to
a continuous-mode region
~\cite{R49_CM1,R50_CM2}. 
Traditional single-mode models are limited in capturing the effects 
introduced by mode variation and cannot adequately describe DSP outputs that 
involve multi-point sampling and processing, 
thereby complicating both the security and performance analysis of the system.

The introduction of temporal modes (TMs) 
~\cite{R51_TM1,R52_TM2,R53_TM3,R54_TM4,R20_CVQKD_T11}
provides a viable approach to resolving the challenges inherent in 
continuous-mode scenarios. 
In this work, we extend this security-analysis framework by constructing 
an entanglement-based (EB) scheme that more accurately captures device nonidealities. 
We further develope a corresponding secret key rate calculation method for 
continuous-mode scenarios. 
Our analysis indicates that the traditional single-mode model arises as 
a special case of our framework under idealized conditions.

We investigate how the receiver's detection bandwidth 
and the transmitter's pulse-shaping formats affect system performance, 
and demonstrate that optimizing the pulse shape can improve performance 
under detector-bandwidth-limited conditions. 
We also verify experimentally that the proposed model accurately captures 
the impact of sampling-time deviations: in our 30-km fiber system, 
a 40-ns offset reduces the secret key rate by 69\%, 
while a 50-ns offset drives it to zero. 
We further introduce a linear weighted-reconstruction DSP method that 
combines multiple sampling points within one pulse period, 
this approach requires no additional hardware and improves the secret key rate by 
approximately 50\% compared with the case without DSP, 
demonstrating a substantial performance enhancement at metropolitan distances.

This paper is organized as follows. 
In Sec.~\ref{II}, we compare the single-mode and continuous-mode scenarios.
In Sec.~\ref{III}, we present numerical simulations based on the proposed model. 
In Sec.~\ref{IV}, we experimentally validate the model. 
Our conclusions are summarized in Sec.~\ref{V. CONCLUSION}.


\begin{figure*}[htb]
    \centering
    \includegraphics[width=\textwidth]{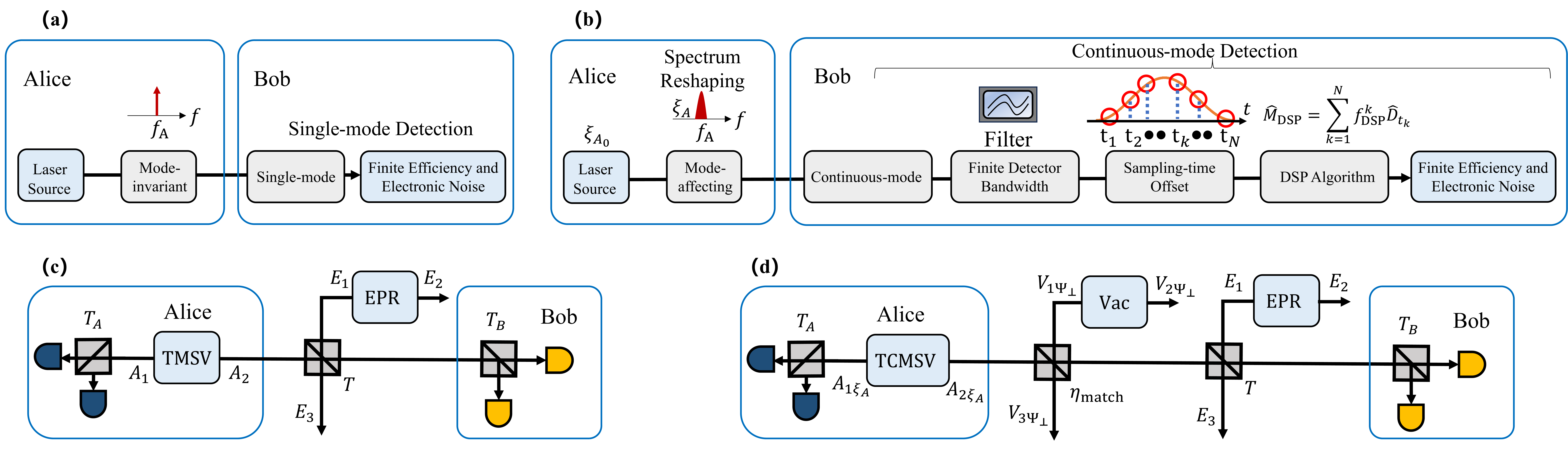}
    \captionsetup{justification=centering, format=plain} 
    \caption{
    \justifying
    Comparison between the single-mode and continuous-mode scenarios.
    (a) PM scheme under the single-mode assumption.
    (b) EB scheme in the continuous-mode scenario.
    (c) EB scheme under the single-mode assumption.
    (d) EB scheme in the continuous-mode scenario.
    When $T_A=1 / 2$, the scheme is equivalent to Alice sending coherent states; 
    when $T_A=1$, it is equivalent to Alice sending squeezed states.
    When $T_B=1 / 2$, the scheme is equivalent to Bob performing heterodyne detection; 
    when $T_B=1$, it corresponds to Bob performing homodyne detection.
    The dark-blue detector represents an ideal detector, 
    while the yellow detector represents a non-ideal detector that includes 
    only detection efficiency and electronic noise.
    $\xi$ denotes a wavepacket that contains time-domain information.
    }
    \label{FIG1}
\end{figure*}

\section{\label{II}Comparison Between the Single-Mode and Continuous-Mode Scenarios}
In this section, we compare the ideal single-mode assumption commonly adopted in 
traditional security analyses with the continuous-mode model that incorporates 
device nonidealities.
For a practical security analysis, the key step is to identify an entanglement-based (EB) 
representation that is equivalent to the corresponding prepare-and-measure (PM) scheme.
In ~\ref{II.A}, we describe the PM scheme of CV-QKD under the single-mode assumption, 
and then introduce the continuous-mode representation that captures device nonidealities.
In ~\ref{II.B}, by analyzing how an ideal single-mode optical field evolves in a practical system, 
we construct the continuous-mode EB scheme.
In our framework, temporal modes (TMs) are employed as an analytical tool for 
conducting security analyses in the continuous-mode  region.
Our results show that the single-mode model emerges as a special case of 
the general continuous-mode framework.

\subsection{\label{II.A}Comparison of Prepare-and-Measure Schemes}
In CV-QKD protocols, the experimental system is typically described using a 
prepare-and-measure (PM) scheme.
Its basic procedure can be summarized in four stages:
First, the transmitter prepares quantum states that carry the encoded information;
then, these quantum states propagate through an untrusted quantum channel;
next, the receiver measures the incoming quantum states;
and finally, the two parties establish a shared secret key through classical post-processing.
The conventional single-mode PM scheme is illustrated in Fig.~\ref{FIG1}(a), 
and its procedure is outlined as follows:

\textbf{1. Preparation of the single-mode quantum state.}
In the state-preparation stage, Alice encodes her information onto the quadrature components of 
a single-mode optical field.

\textbf{2. Transmission of the single-mode quantum state.}
In the transmission stage,
Alice sends the prepared single-mode quantum states to Bob, 
and these states are assumed to preserve their single-mode nature throughout the channel 
without undergoing any mode variation.
An eavesdropper, Eve, may be present in the channel.
Eve may introduce an ancillary quantum system and perform a joint interaction with the 
transmitted states, forwarding part of the signal to Bob while retaining another part 
for her own measurement in an attempt to extract information about the key.

\textbf{3. Measurement of the single-mode quantum state.}
In the measurement stage,
Bob performs non-orthogonal measurements on the received states. 
Using the measurement outcomes, he estimates the channel parameters, 
detects potential eavesdropping attempts, and evaluates both the security and 
the transmission quality of the link.

\textbf{4. Post-processing.}
Finally, Alice and Bob perform classical post-processing, 
including error correction and privacy amplification.

However, the above single-mode PM scheme still relies on several idealized assumptions.
To make the description more consistent with practical conditions, 
we construct the continuous-mode PM scheme, as illustrated in Fig.~\ref{FIG1}(b).
Its procedure is described as follows:

\textbf{1. Preparation of the continuous-mode quantum state.}
Due to nonidealities such as phase noise and spectral broadening 
in the practical laser source,
Alice's light source is no longer an ideal single-mode optical field 
but rather a continuous-mode field,
whose time-domain structure can be characterized by a wavepacket $\xi_{\mathrm{A_0}}$.
Moreover, under the action of the pulse-shaping format, 
this wavepacket further evolves into $\xi_{\mathrm{A}}$, 
which depends on the specific modulation format.
Alice encodes her information onto the quadrature components of 
this continuous-mode optical field.

\textbf{2. Transmission of the continuous-mode quantum state.}
Due to linear and nonlinear effects in the practical channel,
the continuous-mode field $A_{\xi_{\mathrm{A}}}$ emitted by Alice will continue to evolve.

\textbf{3. Measurement of the continuous-mode quantum state.}
Because practical detectors have limited bandwidth,
the continuous-mode states arriving at Bob is effectively subjected to a 
low-pass filtering operation.
Since a continuous-mode state possesses a non-uniform temporal wavepacket, 
sampling at different time leads to different measurement outcomes.
Bob may employ DSP techniques to perform multi-point sampling within one pulse period 
and process the resulting data.
He then uses these measurement outcomes to estimate the channel parameters, 
detect potential eavesdropping attempts, and evaluate both the security and 
the transmission quality of the link.

\textbf{4. Post-processing.}
This step remains unchanged from the single-mode scenario.

The comparison of the single-mode PM scheme and its continuous-mode counterpart shows that 
the continuous-mode PM scheme incorporates a broader range of practical device nonidealities. 
When the experimental devices operate under ideal conditions, 
the continuous-mode model reduces to the single-mode model.

\subsection{\label{II.B}Comparison of Entanglement-Based Schemes}
We perform the security analysis by constructing the entanglement-based (EB) scheme 
that is equivalent to the corresponding PM description.

\textbf{Single-mode EB scheme.}
The single-mode EB scheme is illustrated in Fig.~\ref{FIG1}(c).
In this scheme, each transmission of a single-mode quantum state is equivalent to 
Alice preparing a two-mode squeezed vacuum (TMSV) state~\cite{R56_TMSV} 
and performing a measurement on 
one of its modes.
The single-mode PM and EB schemes generate identical quantum states at the input of 
the quantum channel,
and the EB scheme facilitates quantitative analysis using the von Neumann entropy.

Conventional security analyses rely on the single-mode assumption, 
where the optical field is treated as having a single frequency component.
An ideal single-mode coherent state can be expressed using the 
annihilation operator $\hat{a}_i$ and the creation operator $\hat{a}_i^{\dagger}$ 
of the single-mode field.
Under this assumption, the TMSV state is generated by applying the two-mode squeezing operator 
to the vacuum state:

\begin{equation}
|\mathrm{TMSV}\rangle=\hat{S}_2(r)|\mathrm{vac}\rangle_{a b},
\end{equation}
where
\begin{equation}
\hat{S}_2(r)=\exp \left[r\left(\hat{a} \hat{b}-\hat{a}^{\dagger} \hat{b}^{\dagger}\right)\right]
\end{equation}
is the two-mode squeezing operator, with the squeezing parameter $r$.

However, due to practical device nonidealities, the spectrum of Alice's light source exhibits 
a finite distribution, and pulse modulation further introduces additional variations.
As a result, the optical field in a practical system deviates from an ideal single-mode description, 
and models based on the single-mode assumption are limited in capturing 
the temporal information contained in the practical quantum states.

Bob's practical measurement outcomes also differ from those predicted by the single-mode model.
Although the conventional single-mode EB scheme accounts for several certain nonidealities, 
such as detection efficiency and non-negligible electronic noise, it still neglects 
the impact of limited detector bandwidth on the temporal information carried by 
the quantum states.
For a continuous-mode optical field with a specific temporal distribution, 
the quadrature components vary with the sampling time, 
which is difficult to capture within the single-mode model.
Moreover, when Bob applies DSP algorithms involving multi-point sampling and processing, 
the resulting outputs cannot be directly related to the single-mode description.

In contrast, the continuous-mode EB scheme provides a more accurate description of 
practical systems than the single-mode model. Its formulation is given as follows:

\textbf{Continuous-mode EB scheme.}
The continuous-mode EB scheme is illustrated in Fig.~\ref{FIG1}(d).
Due to the finite linewidth and phase noise of practical lasers, 
the emitted optical field can be regarded as a continuous-mode coherent state. 
When modulation is applied, new frequency components are introduced in the spectral domain, 
and these components become increasingly dense as the modulation speed increases. 
In such situations, a single-mode field operator is no longer sufficient 
to characterize quantum states that possess specific temporal structures.
This motivates the use of continuous-mode field operators to describe these states.
By transforming the discrete-mode operators, the continuous-mode annihilation 
and creation operators can be defined as~\cite{R49_CM1}
$\hat{a}_i \rightarrow \sqrt{\Delta\omega}\hat{a}(\omega)$ and
$\hat{a}_i^{\dagger} \rightarrow \sqrt{\Delta\omega}\hat{a}^{\dagger}(\omega)$,
where $\Delta\omega$ denotes the mode spacing.
The corresponding continuous-mode operators satisfy the commutation relation
$\left[\hat{a}(\omega), \hat{a}^{\dagger}\left(\omega^{\prime}\right)\right] = \delta(\omega - \omega^{\prime})$.

Similar to the TMSV state in the single-mode scenario,
the two-continuous-mode squeezed vacuum (TCMSV) state can be defined by 
the two-continuous-mode squeezing operator acting on the vacuum state:

\begin{equation}
|\mathrm{TCMSV}\rangle=\hat{S}_2(\beta)|\mathrm{vac}\rangle_{a b},
\end{equation}
where
\begin{equation}
\hat{S}_2(\beta)=\exp \left(\hat{P}_{a b}(\beta)-\hat{P}_{a b}^{\dagger}(\beta)\right),
\end{equation}

\begin{equation}
\hat{P}_{a b}^{\dagger}(\beta)=\int d \omega \int d \omega^{\prime} \beta\left(\omega, \omega^{\prime}\right) \hat{a}^{\dagger}(\omega) \hat{b}^{\dagger}\left(\omega^{\prime}\right).
\end{equation}
Here, $\hat{S}_2(\beta)$ denotes the two-continuous-mode squeezing operator, 
and $\beta(\omega, \omega^{\prime})$ represents the two-photon spectrum.

To further analyze continuous-mode states in the time domain,
we take the creation operator as an example and apply the inverse Fourier transform, namely
$\hat{a}^{\dagger}(t)=({1}/{\sqrt{2 \pi}}) \int d \omega \hat{a}^{\dagger}(\omega) \exp (-i \omega t)$.
By defining a temporal wavepacket $\xi_i(t)$, 
the corresponding photon wavepacket creation operator~\cite{R55_COperator} 
can then be written as:
\begin{equation}
\hat{A}_{\xi_i}^{\dagger}=\int d t \xi_i(t) \hat{a}^{\dagger}(t).
\end{equation}
The annihilation operator $\hat{A}_{\xi_i}$ is defined in a similar manner.
When $\xi_i(t)$ satisfies the orthonormal condition
$\left[\hat{A}_{\xi_i}, \hat{A}_{\xi_j}^{\dagger}\right]=\delta_{i j}$,
the operators $\hat{A}_{\xi_i}^{\dagger}$ and $\hat{A}_{\xi_i}$ are 
referred to as TM field operators~\cite{R53_TM3}.
When $\hat{A}_{\xi_{\mathrm{A}}}^{\dagger}$ acts on the vacuum state, 
it generates a coherent state with temporal wavepacket $\xi_\mathrm{A}(t)$.
Unlike the single-mode assumption, in the continuous-mode EB scheme, 
as illustrated in Fig.~\ref{FIG1}(d),
measuring one mode of the TCMSV state is equivalent to 
preparing a continuous-mode quantum state at the transmitter.

After established the equivalence between the EB and PM schemes at the transmitter in the 
continuous-mode scenario,
we now turn to the detection model for continuous-mode quantum states at the receiver.

In practical experiments, owing to the nonidealities of various devices, 
the signals received by Bob typically exhibit a certain temporal distribution.
The quantum state transmitted through the channel can be expressed as
$\left|x_A+i p_A\right\rangle_{\xi_\mathrm{A}}$,
where $\xi_\mathrm{A}$ denotes the wavepacket that carries 
the temporal information~\cite{R20_CVQKD_T11}.
While ideal detection corresponds to measuring the quadrature components of 
the entire wavepacket,
a practical detector can only filter, detect, sample, 
and apply DSP processing to a portion of the wavepacket.

A detailed description of the receiver's modes is provided in Appendix~\ref{Appendix A}. 
After shot-noise calibration and normalization, the creation operator of the receiver TM can be defined as:
\begin{equation}
\hat{A}_{\Xi_{\mathrm{DSP}}}^{\dagger}=\int d \tau \Xi_{\mathrm{DSP}}(\tau) \hat{a}^{\dagger}(\tau),
\end{equation}
$\Xi_{\mathrm{DSP}}(\tau)$ denotes the normalized temporal wavepacket at 
the receiver after DSP processing.

The TM of the quantum state to be measured is denoted by $\xi_{\mathrm{A}}\text{-TM}$, 
and the TM of the receiver is denoted by $\Xi_{\mathrm{DSP}}\text{-TM}$.
Within the TM representation, the continuous-mode detection process can be regarded as 
projecting the state's TM onto the receiver's TM~\cite{R20_CVQKD_T11}.
By applying Gram-Schmidt orthogonalization, a third TM, $\Psi_{\perp}\text{-TM}$, 
can be introduced, which is derived from $\Xi_{\mathrm{DSP}}\text{-TM}$ and orthogonal to $\xi_A\text{-TM}$.
The corresponding decomposition of the creation operator is then given by:

\begin{equation}
    \hat{A}_{\Xi_{\mathrm{DSP}}}^{\dagger}=\sqrt{\eta_{\mathrm{match}}} \hat{A}_{\xi_{\mathrm{A}}}^{\dagger}+\sqrt{1-\eta_{\mathrm{match}}} \hat{A}_{\Psi_{\perp}}^{\dagger},    
\end{equation}
where $\eta_{\text{match}}$ denotes the mode-matching coefficient,
\begin{equation}
    \label{match}
    \eta_{\mathrm{match}}=\left|\int d t \Xi_{\mathrm{DSP}}^*(t) \xi_{\mathrm{A}}(t)\right|^2.
\end{equation}

In summary, Table~\ref{contrast} presents the differences between the single-mode 
and continuous-mode scenarios.

Under the single-mode assumption, an ideal detector is able to extract the complete 
information from an optical field regardless of its temporal wavepacket, 
achieving 100\% detection efficiency in all scenarios.
In practical systems, however, the mode-matching coefficient between the transmitter's TM 
and the receiver's TM significantly affect the detection performance.
For example, the finite bandwidth of Bob's detector and the modulation format applied to 
Alice's pulse wavepacket both influence the final measurement outcome, 
and we will further analyze these effects in Sec.~\ref{III}.
Moreover, supported by experimental results, 
Sec.~\ref{IV} demonstrates how sampling-time deviations and DSP processing affect 
the mode-matching coefficient, thereby impacting the overall system performance.

\begin{table*}[htbp]
    \caption{\label{contrast}%
    Comparison between Single-Mode and Continuous-Mode Scenarios.
    }
    \centering
    \begin{ruledtabular}
    \begin{tabular}{ccc}
    \textnormal{} & \textbf{Single-mode scenarios} & \textbf{Continuous-mode scenarios} \\    \colrule
    \textnormal{Quantum state} & \textnormal{Ideal coherent state $|\alpha\rangle$} & \textnormal{Photon-wavepacket coherent state $|\alpha\rangle_{\xi_{\mathrm{A}}}$} \\
    \textnormal{Spectral characteristics} & \textnormal{Single-frequency spectrum} & \textnormal{Specific spectral distribution} \\
    \textnormal{Temporal characteristics} & \textnormal{No temporal waveform evolution} & \textnormal{Contains temporal-wavepacket information} \\
    \textnormal{Detector} & \textnormal{Fully captures spectral/temporal information of the field} & \textnormal{Bandwidth-limited} \\
    \textnormal{Sampling} & \textnormal{Sampling result invariant with time} & \textnormal{Affected by sampling-time offsets} \\
    \textnormal{Signal processing} & \textnormal{Hard to match DSP} & \textnormal{Matches multi-point processing DSP} \\
    \end{tabular}
    \end{ruledtabular}
\end{table*}

\section{\label{III}Numerical Simulation}

\begin{figure*}[htb]
    \centering
    \includegraphics[width=\textwidth]{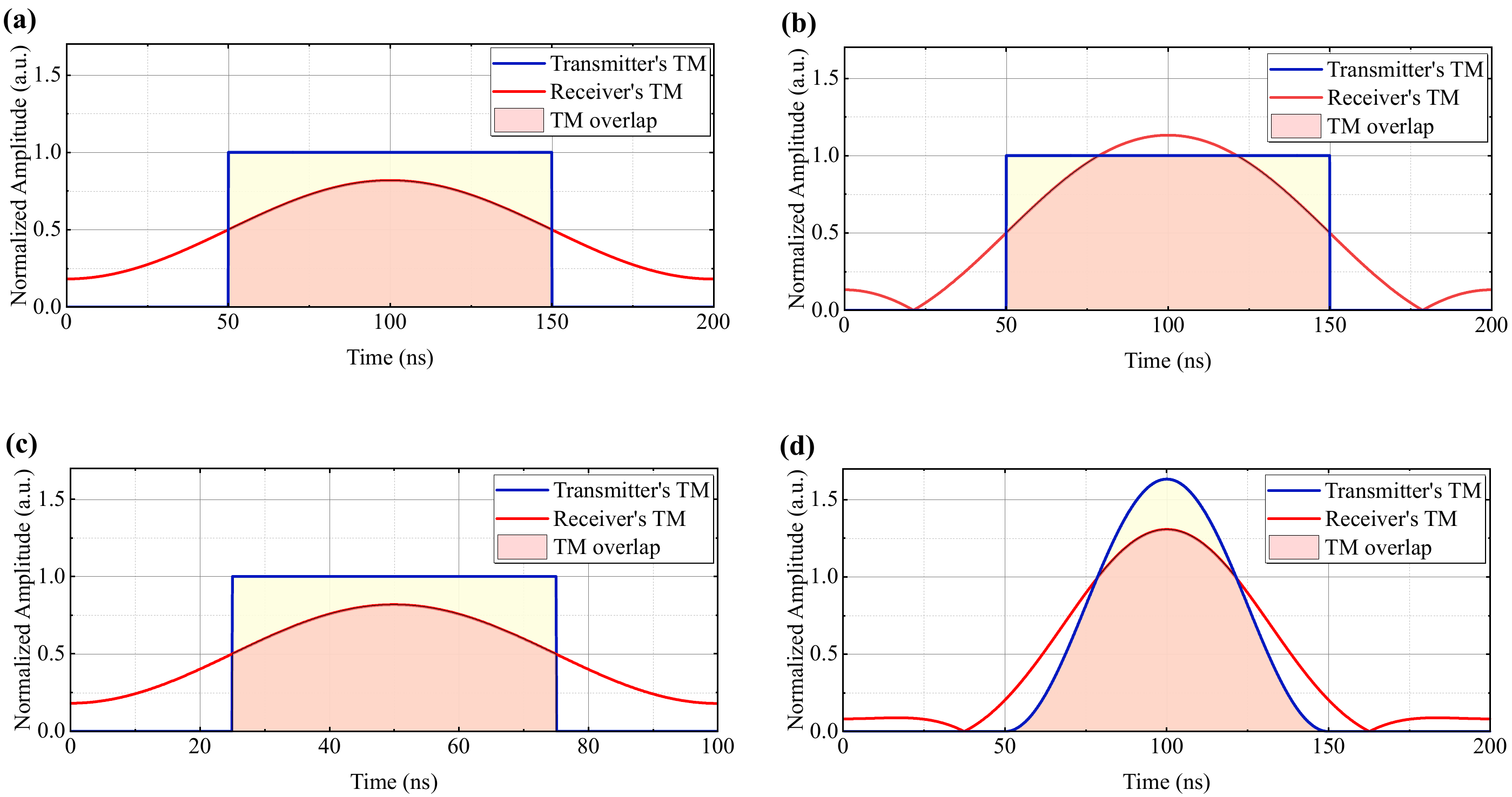}
    \captionsetup{justification=centering, format=plain} 
    \caption{
    \justifying 
    The impact of different detector bandwidths and pulse-shaping formats on 
    the TM matching coefficient between the transmitter and receiver.
    (a) Detector bandwidth: 5-MHz; square pulse with a 5-MHz modulation rate at the transmitter.
    (b) Detector bandwidth: 10-MHz; square pulse with a 5-MHz modulation rate at the transmitter.
    (c) Detector bandwidth: 10-MHz; square pulse with a 10-MHz modulation rate at the transmitter.
    (d) Detector bandwidth: 10-MHz; raised-cosine pulse with a 5-MHz modulation rate at the transmitter.
    }
    \label{FIG2}
\end{figure*}

The main distinction between continuous-mode and single-mode quantum states is that 
continuous-mode states carry the spectral/temporal distribution of the optical field. 
We characterize this structure using temporal modes (TMs). 
Since each TM behaves as an effective single mode, 
the corresponding security analysis can be carried out in the single-mode framework. 
After establishing the PM-EB equivalence, the security analysis proceeds directly within 
the EB scheme.

For clarity of analysis, we focus on the secret key rate under collective attacks 
in the asymptotic regime. The finite-size correction terms do not affect the core part of 
our model. The secret key rate with reverse reconciliation is given by~\cite{R57_KeyRate1,R58_KeyRate2}:
\begin{equation}
K=\beta_\mathrm{R} I(A: B)-\chi(B: E),
\end{equation}
where $\beta_\mathrm{R}$ denotes the reconciliation efficiency, 
$I(A: B)$is the mutual information between Alice and Bob, 
and $\chi(B: E)$ is the Holevo bound between Bob and Eve.
In the experiment, Alice and Bob obtain the covariance matrix $\mathrm{V}_{\mathrm{AB}}$ 
through the parameter estimation procedure, from which
$I(A: B)$ and $\chi(B: E)$ are calculated. The covariance matrix is given by:

\begin{equation}
\mathbf{V}_{\mathrm{AB}}=\left(\begin{array}{cc}
a \mathbf{I} & c \mathbf{Z} \\
c \mathbf{Z} & b \mathbf{I}
\end{array}\right).
\end{equation}
Where $a=V_{\mathrm{A}}+1$,$V_{\mathrm{A}}$represents the modulation variance at the Alice's side.
$c=\sqrt{\eta_{\mathrm{tot}}\left(V^2-1\right)}$.$V=V_{\mathrm{A}}+1$.
$b=\eta_{\mathrm{tot}}\left(V+\chi_{\mathrm{match}}+\chi_{\mathrm{C}}+\chi_{\mathrm{D}}\right)$,
$I=\mathrm{diag}(1, 1)$, $Z=\mathrm{diag}(1, -1)$. And

\begin{equation}
\eta_{\mathrm{tot}}=\eta_{\mathrm{C}} \eta_{\mathrm{match}} \eta_{\mathrm{D}},
\end{equation}

\begin{equation}
\chi_{\mathrm{match}}=\frac{1-\eta_{\mathrm{match}}}{\eta_{\mathrm{match}}},
\end{equation}

\begin{equation}
\chi_{\mathrm{C}}=\frac{1}{\eta_{\mathrm{match}}}\left(\frac{1-\eta_{\mathrm{C}}}{\eta_{\mathrm{C}}}+\varepsilon\right),
\end{equation}

\begin{equation}
\chi_{\mathrm{D}}=\frac{1-\eta_{\mathrm{D}}+v_{\mathrm{el}}}{\eta_{\mathrm{C}} \eta_{\mathrm{match}} \eta_{\mathrm{D}}}.
\end{equation}

In the trusted-detection model~\cite{R59_DetectM}, 
$\eta_{\text {tot}}$ denotes the total loss, 
and $\eta_{\mathrm{C}}=10^{-\alpha L / 10}$ represents the channel transmittance, 
assuming a channel loss of $\alpha=0.2 \mathrm{~dB} / \mathrm{km}$ and $L$ 
is the transmission distance. $\eta_{\text {match }}$ characterizes 
the loss induced by TM mismatch, and $\eta_{\mathrm{D}}$ denotes 
the detection efficiency. $\varepsilon$ represents the excess noise, 
while $v_{\mathrm{el}}$ corresponds to the electronic noise.
In this section, we consider the coherent state and homodyne detection protocol as 
the simulation model, whose EB representation is shown in Fig.~\ref{FIG1}(d), 
where $T_A=1 / 2$ and $T_B=1$. The parameters used in the simulation are: 
$V_{\mathrm{A}}=4$, $\beta_\mathrm{R}=0.95$, $L$=10-km, $\varepsilon=0.01$, 
$\eta_{\mathrm{D}}=0.481$, and $v_{\mathrm{el}}=0.0584$.

We use two examples to show how device nonidealities in practical systems alter the 
mode-matching coefficient $\eta_{\text{match}}$ and consequently affect the secret key rate. 
The detailed calculation method is provided in Appendix~\ref{Appendix B}.

\begin{figure*}[htb]
    \centering
    \includegraphics[width=\textwidth]{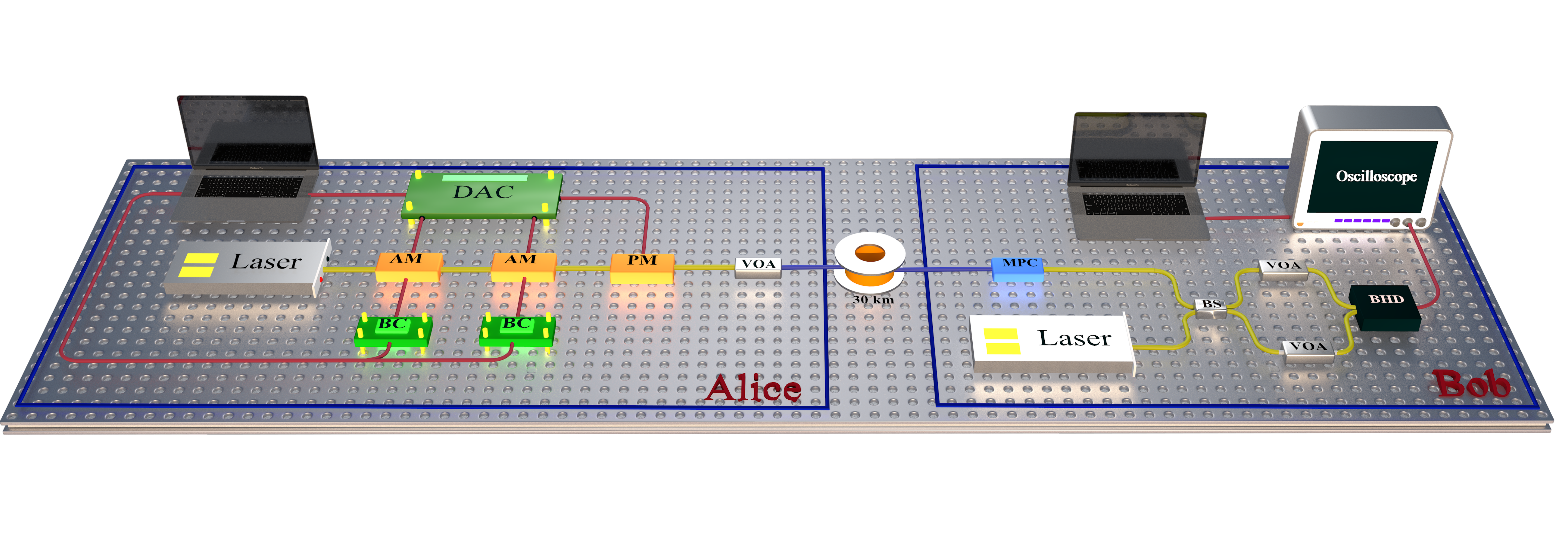}
    \captionsetup{justification=centering, format=plain} 
    \caption{
    \justifying 
    Experimental optical setup for the CV-QKD system.
    We perform a Gaussian-modulated coherent state and homodyne detection experiment. 
    We analyze the impact of sampling-time offsets on the secret key rate based on 
    the proposed theoretical model. We also evaluate the performance improvement 
    provided by the linear weighted-reconstruction DSP method. 
    AM: amplitude modulator. PM: phase modulator. DAC: digital-to-analog converter. 
    BC: bias controller. VOA: variable optical attenuator. MPC: manual polarization controller. 
    BS: beam splitter. BHD: balanced homodyne detector.
    }
    \label{FIG3}
\end{figure*}

\textbf{1. Detector bandwidth.}
An ideal detector has an infinite bandwidth and can extract all information from 
optical pulses with arbitrary temporal wavepacket, which leads to $\eta_{\text{match}}$ of 
100\% at any sampling time. In practical systems, however, the detection bandwidth at 
the receiver is always limited, which affects the degree of mode matching between 
the transmitter's TM and the receiver's TM, consequently influences the system performance. 
In addition, the modulation applied at the transmitter inevitably broadens the pulse spectrum, 
making it difficult for the receiver to completely reconstruct the information contained in 
the original wavepacket.
Since $\eta_{\text{match}}$ is jointly determined by multiple factors, 
in the simulations of this section we assume perfect sampling and ideal digital compensation algorithm. 
We vary only the detection bandwidth in order to clearly evaluate its impact on $\eta_{\text{match}}$. 
In Fig.~\ref{FIG2}, the transmitter's TM is shown as the blue curve and 
the receiver's TM is shown as the red curve.

In Fig.~\ref{FIG2}(a), a 5-MHz square pulse with a 50\% duty cycle is modulated 
at the transmitter and detected by a 5-MHz-bandwidth detector at the receiver. 
According to Eq.~\ref{match}, the $\eta_{\text{match}}$ between the transmitter's TM 
and receiver's TM is 0.822.
To improve system performance, Fig.~\ref{FIG2}(b) keeps the same 5-MHz square pulse 
at the transmitter while increasing the detector bandwidth to 10-MHz, 
which raises $\eta_{\text{match}}$ to 0.907.
To analyze the effect of spectral broadening caused by increasing the modulation rate, 
Fig.~\ref{FIG2}(c) keeps the 10-MHz detector bandwidth at the receiver 
but raises the modulation repetition rate to 10-MHz, 
leading to a reduction of $\eta_{\text{match}}$ back to 0.822.

\textbf{2. Pulse-shaping formats.}
Because of the limited detection bandwidth, modifying the pulse-shaping format at 
the transmitter can be used to improve system performance. For example, 
replacing the square pulse with a raised-cosine (RC) pulse yields lower spectral sidelobes 
and concentrates more energy in the main lobe, thereby achieving higher waveform fidelity 
and a higher $\eta_{\text{match}}$ under detector-bandwidth-limited conditions. 
We perform the following simulation: in Fig.~\ref{FIG2}(d), 
the receiver still employs a detector with a 10-MHz bandwidth, 
while the 5-MHz square pulse at the transmitter is replaced by a 5-MHz RC pulse. 
As the RC pulse is better suited for detector-bandwidth-limited transmission, 
$\eta_{\text{match}}$ increases to 0.941 compared with the square-pulse case.

The impact of $\eta_{\text{match}}$ on the secret key rate is summarized in 
Table~\ref{tableII}, where BW denotes the bandwidth.

\begin{table}[h!]
    \centering
    \caption{Impact of different mode-matching coefficients on the secret key rate. RC: raised-cosine.}
    \begin{ruledtabular}
    \begin{tabular}{ccc}  
    \textbf{Condition} & \textbf{$\eta_{\text {match}}$} & \textbf{Secret key rate } \\
    \colrule
    Ideal scenario & 1 & 0.20 (bit/pulse) \\
   Square; 5-MHz detector BW & 0.822 & 0.04 (bit/pulse)\\
   Square; 10-MHz detector BW & 0.907 & 0.12 (bit/pulse)\\
   RC; 10-MHz detector BW & 0.941 & 0.15 (bit/pulse)\\
    \end{tabular}
    \end{ruledtabular}
    \label{tableII}
\end{table}

The detector bandwidth at the receiver and the modulation rate at the transmitter 
both have a significant impact on the TM matching between the two sides. 
Increasing the detection bandwidth can partially compensate for the mismatch 
induced by spectral broadening. However, when the modulation rate becomes higher, 
the spectrum broadens further and $\eta_{\text{match}}$ decreases again.
In addition, optimizing the pulse modulation format can improve $\eta_{\text{match}}$ between 
the transmitter's TM and the receiver's TM, 
which provides better performance in detector-bandwidth-limited scenarios. 
For future systems, TM matching should be considered as 
one of the key parameters in system design.

\section{\label{IV}Experiments and Analysis}
In this section, we conduct an experiment with the coherent state and homodyne detection, 
then we use our continuous-mode model to analyze how sampling-time offsets influence 
the secret key rate. We also demonstrate that DSP algorithm can improve the system performance. 
The experimental setup is shown in Fig.~\ref{FIG3}, and our configuration 
is summarized as follows.

At the transmitter, a narrow-linewidth laser serves as the optical source. 
A 5MHz square-wave pulse train is generated by the first amplitude modulator (AM), 
corresponding to a pulse duration of 200ns. To clearly observe the wavepacket variations, 
the duty cycle is set to 50\%. The square pulses then pass through an AM 
and a phase modulator (PM) to realize Gaussian modulation, 
after which the modulated coherent states are transmitted through a 30-km fiber channel.

At the receiver, another narrow-linewidth laser generates a continuous-wave local oscillator, 
which is combined with the incoming signal on a 50:50 beam splitter (BS) 
and measured using a balanced homodyne detector. 
To mitigate the bandwidth-induced mode mismatch discussed in Sec.~\ref{III}, 
we employ a detector with a bandwidth of 1.6-GHz, which is much larger than 
the 5-MHz pulse modulation rate. The detector output is recorded by an oscilloscope operating 
at a sampling rate of 500-MSa/s, resulting in 100 sampling points for each transmitted pulse.

In the single-mode assumption, the optical field is temporally invariant, 
and therefore sampling at different time yields identical information.

\begin{figure}[h]
    \centering
    \includegraphics[width=0.48\textwidth]{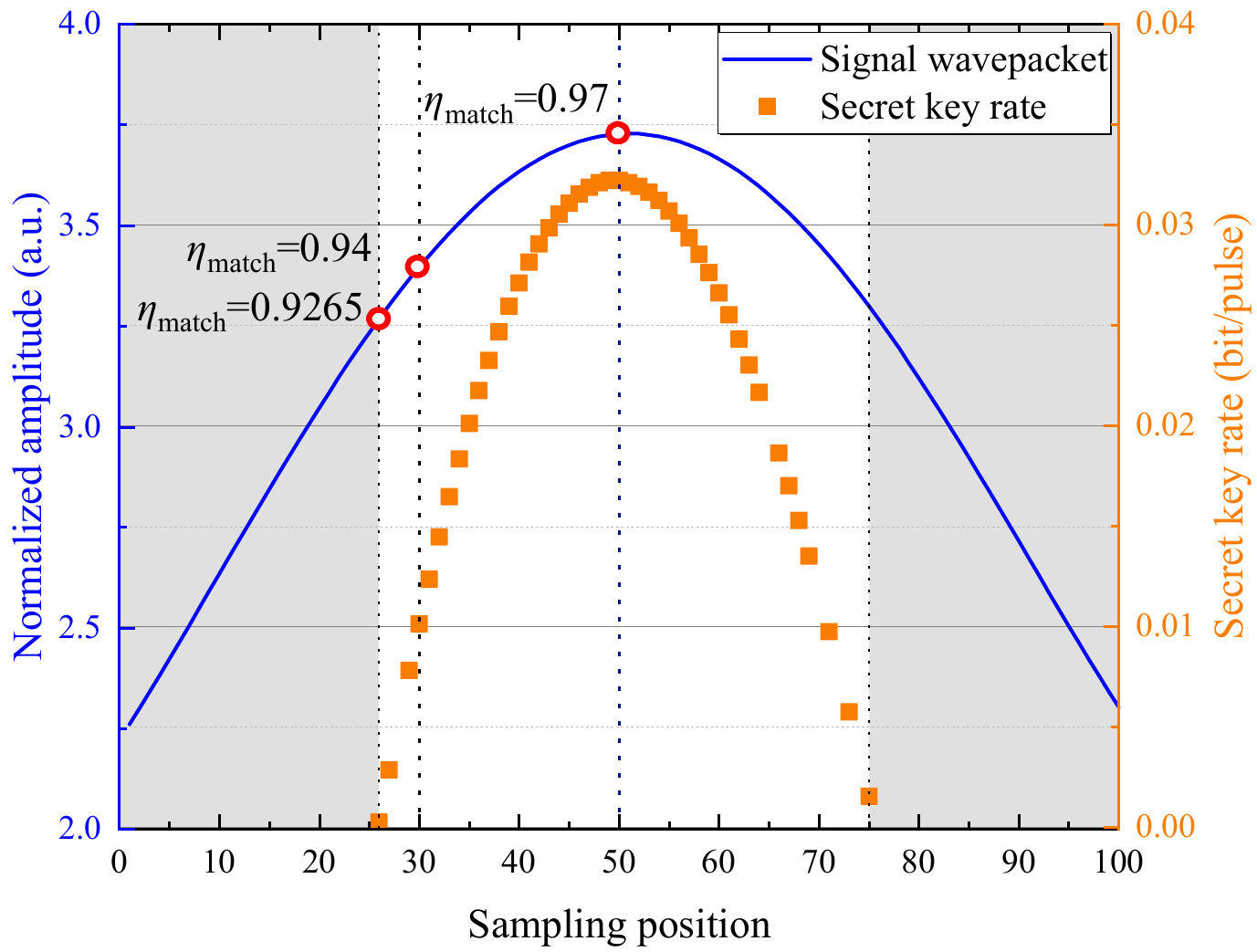}
    \caption{
    \justifying 
    Experimental results for the 30-km fiber experiment.
    The blue curve represents the wavepacket reconstructed from the receiver data.
    The orange markers indicate the secret key rates obtained at different sampling points.
    The gray region corresponds to zero key rate.
    }
    \label{FIG4}
\end{figure}

However, due to various nonidealities in practical systems, 
such as pulse broadening, dispersion, and imperfections in the compensation algorithms, 
the signal wavepacket evolves over time. 
In this case, the sampling time directly affects $\eta_{\text{match}}$ between 
the transmitter's TM and receiver's TM, 
thereby influencing the system performance. 
The experimental results are shown in Fig.~\ref{FIG4}. 
The parameters used in the experiment are as follows:
$V_{\mathrm{A}} = 3.9892$, $\beta_{\mathrm{R}} = 0.95$, $L = 30~\mathrm{km}$, 
$\varepsilon = 0.01179$, $\eta_{\mathrm{D}} = 0.481$, and $v_{\mathrm{el}} = 0.0584$.

As shown in Fig.~\ref{FIG4}, when the sampling time is close to the center of the pulse, 
the TM-matching coefficient $\eta_{\text{match}}$ reaches 0.97, 
yielding the highest secret key rate of 0.032 bit/pulse under single-point sampling. 
As the sampling time deviates from the pulse center, $\eta_{\text{match}}$ gradually decreases, 
leading to a corresponding reduction in the secret key rate. 
At the 30th sampling point (corresponding to a 40-ns timing offset), 
$\eta_{\text{match}}$ drops to 0.94, and the secret key rate decreases to 0.010 bit/pulse. 
The lowest secret key rate occurs at the 26th sampling point 
(corresponding to a 48-ns timing offset), where $\eta_{\text{match}}$ further 
decreases to 0.9265 and the key rate falls to 0.000265 bit/pulse, approaching zero.

In the single-point sampling scheme, even when the secret key rate reaches its maximum, 
the TM-matching coefficient $\eta_{\text{match}}$ at the corresponding sampling point 
still has room for improvement, indicating that the system performance can be further optimized. 
Since the signal wavepacket remains relatively stable only near the center of each period 
and exhibits noticeable distortion toward the edges, we select 30 sampling points 
(from the 36th to the 65th) as the effective sampling region for each pulse. 
The samples within this region are linearly weighted and averaged to reconstruct the final data 
of each pulse. Using this method, $\eta_{\text{match}}$ is improved from 0.97 
(under single-point sampling) to 0.995, and the secret key rate increases to 0.049 bit/pulse, 
corresponding to an approximately 50\% enhancement without requiring any additional hardware.

\begin{figure}[h]
    \centering
    \includegraphics[width=0.48\textwidth]{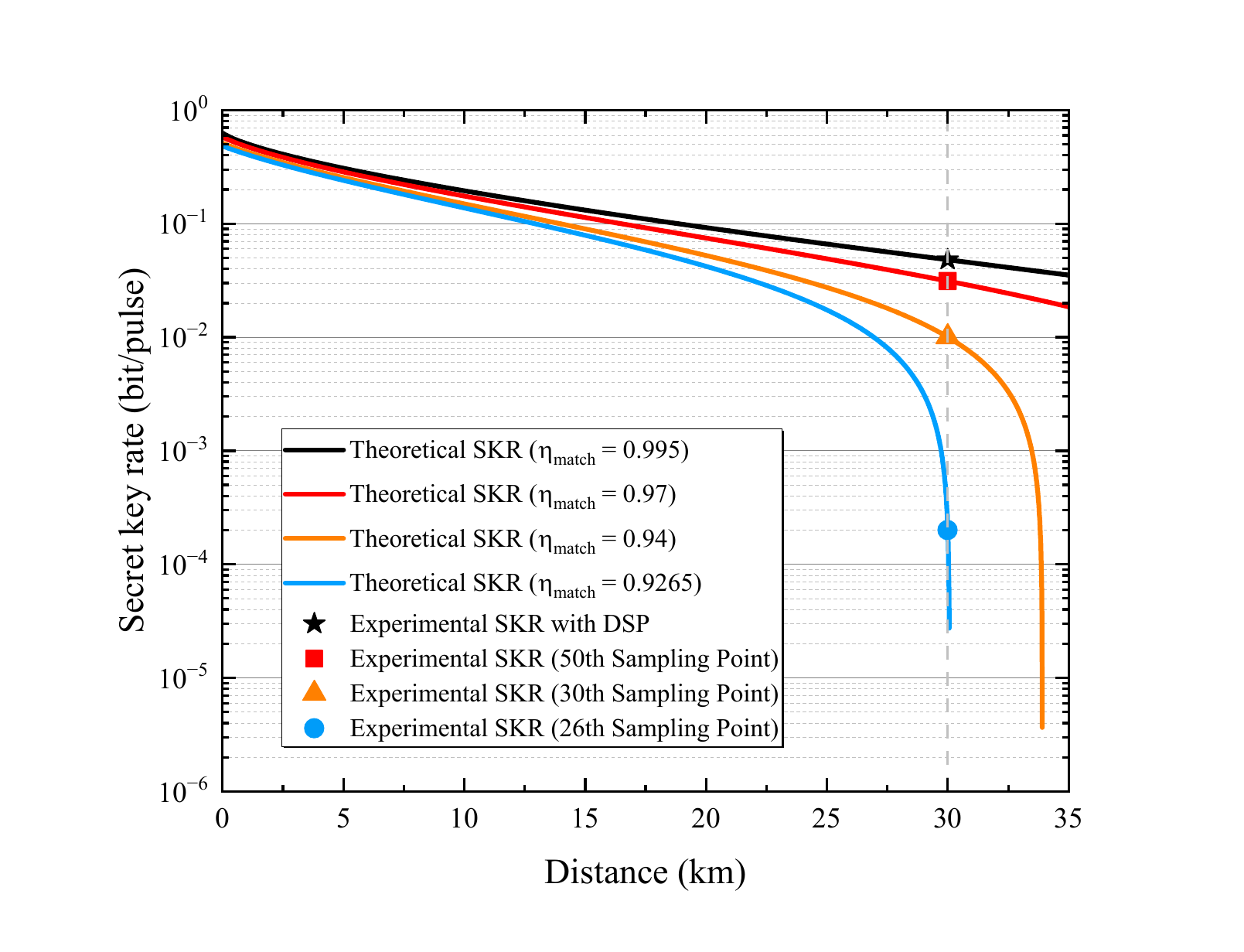}
    \caption{
    \justifying 
   Simulation of the impact of different mode-matching coefficients on system performance.
    }
    \label{FIG5}
\end{figure}

The simulation and experimental results are shown in Fig.~\ref{FIG5}.
The black curve represents the theoretical secret key rate with DSP, 
where $\eta_{\text{match}}$ is improved to 0.995.
The red, orange, and blue curves correspond to the theoretical secret key rates for 
$\eta_{\text{match}} = 0.97$, $\eta_{\text{match}} = 0.94$, and $\eta_{\text{match}} = 0.9265$, 
respectively.
The star markers denote the experimental results obtained using the DSP method.
The square markers indicate the experimental results obtained by sampling at the 
50th point of the pulse, which lies at the pulse center.
The triangle markers correspond to sampling at the 30th point, 
where the sampling-time offset is 40-ns.
The circular markers correspond to sampling at the 26th point, 
where the sampling-time offset is 48-ns.

The sampling-time offset affects $\eta_{\text{match}}$ between the transmitter's TM 
and receiver's TM, thus impacts the secret key rate. 
The linear DSP method can partially compensate for this mode mismatch and 
consequently improve the system performance.

\section{\label{V. CONCLUSION}CONCLUSION}
In this work, we conducted a security and performance analysis of CV-QKD in 
continuous-mode scenarios that more closely reflect practical systems. 
By introducing TMs, we established an EB scheme capable of 
characterizing various experimental nonidealities, 
such as finite detection bandwidth at the receiver, 
wavepacket evolution of the transmitted signal, sampling-time offsets, 
and DSP algorithm. These effects are quantified within a unified framework through the 
mode-matching coefficient $\eta_{\text{match}}$. 
The model also includes the conventional single-mode assumption as a special case, 
where $\eta_{\text{match}}$ takes the specific value of 1.

We present a concrete method for calculating the secret key rate in the continuous-mode scenario. 
We demonstrate that the detection bandwidth, modulation rate, 
and pulse-shaping format all influence the TM matching between the transmitter and receiver, 
thereby affecting the system performance. Under bandwidth-limited conditions, 
increasing the detector bandwidth or optimizing the pulse-shaping format can 
mitigate the mismatch caused by spectral broadening.

We analyze the impact of sampling-time deviations on $\eta_{\text{match}}$ 
and system performance through experiment. A timing offset reduces $\eta_{\text{match}}$, 
and in our 30-km fiber links, the secret key rate drops to zero when the offset reaches 50-ns. 
By introducing a DSP method to improve $\eta_{\text{match}}$ between the transmitter's TM 
and receiver's TM, we achieve an approximately 50\% performance enhancement 
without requiring any additional hardware. 
These results demonstrate that digital compensation can significantly 
improve system performance at metropolitan distances and offers 
further potential for optimization.

The analytical framework developed in this work can accommodate a broader range of 
experimental conditions and provides theoretical guidance for performance improvements 
in digitally implemented CV-QKD systems.

\section*{Acknowledgments}
This work was supported by the National Natural Science Foundation of China 
under Grant No.62371060, No.62201012, No.62001041, NO.62571006, 
the Fund of State Key Laboratory of Information Photonics 
and Optical Communications under Grant No. IPOC2022ZT09. 

%

\appendix
\section{\label{Appendix A}Measurement of continuous-mode quantum states}
In the single-mode scenario, the photocurrent flux operator of the homodyne detector 
is given by:
\begin{equation}
\hat{f}=\left[\hat{a}^{\dagger} \hat{a}_{\mathrm{LO}}+\hat{a}_{\mathrm{LO}}^{\dagger} \hat{a}\right].
\end{equation}

However, the single-mode model cannot adequately capture the temporal information contained 
in practical quantum states. By introducing the continuous-mode operators 
$\hat{a}^{\dagger}(t)$ and $\hat{a}(t)$, and modeling the detector's finite bandwidth 
as a filter with an impulse response function (IRF) $g(t)$, the photocurrent flux operator 
of the homodyne detector in the continuous-mode scenario can be written as:
\begin{equation}
\hat{f}(t)=\left[\hat{a}^{\dagger}(t) \hat{a}_{\mathrm{LO}}(t)+\hat{a}_{\mathrm{LO}}^{\dagger}(t) \hat{a}(t)\right] * g(t),
\end{equation}
where $*$ denotes the convolution.

The photon wavepacket of the local oscillator (LO) is given by:
\begin{equation}
\alpha_{\mathrm{LO}}(t)=\mu_{\mathrm{LO}}^{1 / 2} \xi_{\mathrm{LO}}(t) \exp \left(-i \omega_{\mathrm{LO}} t+i \theta+i \Delta \varphi_{\mathrm{LO}}(t)\right),
\end{equation}
where $\mu_{\mathrm{LO}}$ denotes the average number of photons contained in the 
wavepacket $\xi_{\mathrm{LO}}(t)$ over the defined time interval. 
For most experiments, the local oscillator has a flat wavepacket, 
and thus one may take $\xi_{\mathrm{LO}}(t) = 1$. 
The parameter $\theta$ represents the phase angle, 
and $\Delta \varphi_{\mathrm{LO}}(t)$ denotes the phase noise of the local oscillator.

The photocurrent flux after taking the average over LO is
\begin{equation}
\begin{aligned}
\hat{f}_{\mathrm{LO}}(t) & =\left\langle\alpha_{\mathrm{LO}}(t)\right| \hat{f}(t)\left|\alpha_{\mathrm{LO}}(t)\right\rangle \\
& =\mu_{\mathrm{LO}}^{1 / 2} X_{\mathrm{LO}}^\theta(t) * g(t),
\end{aligned}
\end{equation}
where
\begin{equation}
\begin{aligned}
X_{\mathrm{LO}}^\theta(t) & =\exp \left(-i \omega_{\mathrm{LO}} t+i \theta+i \Delta \varphi_{\mathrm{LO}}(t)\right) \hat{a}^{\dagger}(t) \\
& +\exp \left(i \omega_{\mathrm{LO}} t-i \theta-i \Delta \varphi_{\mathrm{LO}}(t)\right) \hat{a}(t).
\end{aligned}
\end{equation}

For a pulse sampled at $N$ points in total, the $k$-th point is sampled at time $t_k$. 
Assuming that the integration time for each sample is sufficiently short and 
the signal remains constant during this interval, the sampling outcome at the receiver 
at time $t_k$ can be expressed as:
\begin{equation}
\hat{D}_{t_k}=\frac{1}{\Delta t_{\mathrm{s}}} \int_{t_k}^{t_k+\Delta t_{\mathrm{s}}} d t \hat{f}_{\mathrm{LO}}(t),
\end{equation}
it can also be expressed as:
\begin{equation}
\hat{D}_{t_k}=\left.\mu_{\mathrm{LO}}^{1 / 2} X_{\mathrm{LO}}^\theta(t) * g(t)\right|_{t=t_k}.
\end{equation}

When a linear DSP algorithm is applied to process the sampling outcomes, 
its output is given by:
\begin{equation}
\begin{aligned}
\hat{M}_{\mathrm{DSP}} & =\sum_{k=1}^N f_{\mathrm{DSP}}^k \hat{D}_{t_k} \\
& =\mu_{\mathrm{LO}}^{1 / 2} \int d \tau X_{\mathrm{LO}}^\theta(\tau) G_{\mathrm{DSP}}(\tau),
\end{aligned}
\end{equation}
where
\begin{equation}
G_{\mathrm{DSP}}(\tau)=\sum_{k=1}^N f_{\mathrm{DSP}}^k g\left(t_k-\tau\right) .
\end{equation}

$\langle 0| \hat{M}_{\mathrm{DSP}}|0\rangle=0$,
$\langle 0| \hat{M}_j \hat{M}_j|0\rangle=\mu_{\mathrm{LO}} \int d \tau\left[G_{\mathrm{DSP}}(\tau)\right]^2$,
the shot-noise variance can be calculated as:
\begin{equation}
\sigma_{\mathrm{SNU}}^2=\mu_{\mathrm{LO}} \int d \tau\left[G_{\mathrm{DSP}}(\tau)\right]^2.
\end{equation}

The normalized photon wavepacket function can be defined as:
\begin{equation}
\Xi_{\mathrm{DSP}}(\tau)=\frac{G_{\mathrm{DSP}}(\tau) \exp \left(-i \omega_{\mathrm{LO}} \tau+i \theta+i \Delta \varphi_{\mathrm{LO}}(\tau)\right)}{\sigma_{\mathrm{cal}}}.
\end{equation}

The photon-wavepacket creation operator can be defined as:
\begin{equation}
\hat{A}_{\Xi_{\mathrm{DSP}}}^{\dagger}=\int d \tau \Xi_{\mathrm{DSP}}(\tau) \hat{a}^{\dagger}(\tau).
\end{equation}

The output of the receiver can be expressed as:
\begin{equation}
\hat{M}_{\mathrm{DSP}}^{\mathrm{SNU}}=\hat{X}^\theta\left(\Xi_{\mathrm{DSP}}\right)=\hat{A}_{\Xi_{\mathrm{DSP}}}^{\dagger}+\hat{A}_{\Xi_{\mathrm{DSP}}}.
\end{equation}

\section{\label{Appendix B}Secret key rate calculation}
Under the trusted detection model,
the secret key rate with reverse reconciliation is given by
\begin{equation}
K=\beta_\mathrm{R} I(A: B)-\chi(B: E),
\end{equation}
where $\beta_\mathrm{R}$ denotes the reconciliation efficiency, 
$I(A: B)$is the mutual information between Alice and Bob, 
and $\chi(B: E)$ is the Holevo bound between Bob and Eve.

\begin{equation}
I(A: B)_{\mathrm{hom}}=\frac{1}{2} \log _2 \frac{V+\chi_{\mathrm{tot}}}{1+\chi_{\mathrm{tot}}},
\end{equation}
$\chi_{\text {tot }}=\chi_{\text {match }}+\chi_{\mathrm{C}}+\chi_{\mathrm{D}}$,
the relevant parameters are defined in Sec.~\ref{III} of the main text.

Since it has been shown that Gaussian attacks are optimal
for collective attacks~\cite{R60_GAttack1,R61_GAttack2}, 
$\chi(B: E)$ can be written as:
\begin{equation}
\chi(B: E)=\sum_{i=1}^2 G\left(\frac{\lambda_i-1}{2}\right)-\sum_{i=3}^5 G\left(\frac{\lambda_i-1}{2}\right),
\end{equation}
where $G(x)=(x+1) \log _2(x+1)-x \log _2 x$.
\begin{equation}
\lambda_{1,2}^2=\frac{1}{2}\left[A \pm \sqrt{A^2-4 B}\right],
\end{equation}
\begin{equation}
A=a^2+b_0^2-2 c_0^2,
\end{equation}
\begin{equation}
B=\left(a b_0-c_0^2\right)^2,
\end{equation}
where $a=V_{\mathrm{A}}+1$,$V_{\mathrm{A}}$represents the modulation variance at the Alice's side. 
$b_0=T \eta_{\text {match }}\left(V+\chi_{\text {match }}+\chi_{\mathrm{C}}\right)$,
$c_0=\sqrt{T \eta_{\text {match }}\left(V^2-1\right)}$,$V=V_{\mathrm{A}}+1$.
the relevant parameters are defined in Sec.~\ref{III} of the main text.

\begin{equation}
\lambda_{3,4}^2=\frac{1}{2}\left[C \pm \sqrt{C^2-4 D}\right],
\end{equation}
where for the homodyne case,
\begin{equation}
C_{\mathrm{hom}}=\frac{A \chi_{\mathrm{hom}}+V \sqrt{B}+b_0}{T\left(V+\chi_{\mathrm{tot}}\right)},
\end{equation}
\begin{equation}
D_{\mathrm{hom}}=\sqrt{B} \frac{V+\sqrt{B} \chi_{\mathrm{hom}}}{T\left(V+\chi_{\mathrm{tot}}\right)},
\end{equation}
where $\chi_{\text {hom }}=\left[\left(1-\eta_{\mathrm{D}}\right)+v_{\mathrm{el}}\right] / \eta_{\mathrm{D}}$,
$\eta_{\mathrm{D}}$ denotes 
the detection efficiency, 
$v_{\mathrm{el}}$ denotes the electronic noise.

The last symplectic eigenvalue is $\lambda_5 = 1$, 
the secret key rate $K$ of the coherent state and homodyne detection protocol 
can be obtained using the above formulas.

\nocite{*}




%


\begin{thebibliography}{61}%
\makeatletter
\providecommand \@ifxundefined [1]{%
 \@ifx{#1\undefined}
}%
\providecommand \@ifnum [1]{%
 \ifnum #1\expandafter \@firstoftwo
 \else \expandafter \@secondoftwo
 \fi
}%
\providecommand \@ifx [1]{%
 \ifx #1\expandafter \@firstoftwo
 \else \expandafter \@secondoftwo
 \fi
}%
\providecommand \natexlab [1]{#1}%
\providecommand \enquote  [1]{``#1''}%
\providecommand \bibnamefont  [1]{#1}%
\providecommand \bibfnamefont [1]{#1}%
\providecommand \citenamefont [1]{#1}%
\providecommand \href@noop [0]{\@secondoftwo}%
\providecommand \href [0]{\begingroup \@sanitize@url \@href}%
\providecommand \@href[1]{\@@startlink{#1}\@@href}%
\providecommand \@@href[1]{\endgroup#1\@@endlink}%
\providecommand \@sanitize@url [0]{\catcode `\\12\catcode `\$12\catcode
  `\&12\catcode `\#12\catcode `\^12\catcode `\_12\catcode `\%12\relax}%
\providecommand \@@startlink[1]{}%
\providecommand \@@endlink[0]{}%
\providecommand \url  [0]{\begingroup\@sanitize@url \@url }%
\providecommand \@url [1]{\endgroup\@href {#1}{\urlprefix }}%
\providecommand \urlprefix  [0]{URL }%
\providecommand \Eprint [0]{\href }%
\providecommand \doibase [0]{https://doi.org/}%
\providecommand \selectlanguage [0]{\@gobble}%
\providecommand \bibinfo  [0]{\@secondoftwo}%
\providecommand \bibfield  [0]{\@secondoftwo}%
\providecommand \translation [1]{[#1]}%
\providecommand \BibitemOpen [0]{}%
\providecommand \bibitemStop [0]{}%
\providecommand \bibitemNoStop [0]{.\EOS\space}%
\providecommand \EOS [0]{\spacefactor3000\relax}%
\providecommand \BibitemShut  [1]{\csname bibitem#1\endcsname}%
\let\auto@bib@innerbib\@empty
\bibitem [{\citenamefont {Bennett}\ and\ \citenamefont
  {Brassard}(2014)}]{R1_QKD1}%
  \BibitemOpen
  \bibfield  {author} {\bibinfo {author} {\bibfnamefont {C.~H.}\ \bibnamefont
  {Bennett}}\ and\ \bibinfo {author} {\bibfnamefont {G.}~\bibnamefont
  {Brassard}},\ }\bibinfo {title} {Quantum cryptography: Public key
  distribution and coin tossing},\ \href@noop {} {\bibfield  {journal}
  {\bibinfo  {journal} {Theoretical computer science}\ }\textbf {\bibinfo
  {volume} {560}},\ \bibinfo {pages} {7} (\bibinfo {year} {2014})}\BibitemShut
  {NoStop}%
\bibitem [{\citenamefont {Ekert}(1991)}]{R2_QKD2}%
  \BibitemOpen
  \bibfield  {author} {\bibinfo {author} {\bibfnamefont {A.~K.}\ \bibnamefont
  {Ekert}},\ }\bibinfo {title} {Quantum cryptography based on bell's theorem},\
  \href@noop {} {\bibfield  {journal} {\bibinfo  {journal} {Phys. Rev. Lett.}\
  }\textbf {\bibinfo {volume} {67}},\ \bibinfo {pages} {661} (\bibinfo {year}
  {1991})}\BibitemShut {NoStop}%
\bibitem [{\citenamefont {Gisin}\ \emph {et~al.}(2002)\citenamefont {Gisin},
  \citenamefont {Ribordy}, \citenamefont {Tittel},\ and\ \citenamefont
  {Zbinden}}]{R3_QKD3}%
  \BibitemOpen
  \bibfield  {author} {\bibinfo {author} {\bibfnamefont {N.}~\bibnamefont
  {Gisin}}, \bibinfo {author} {\bibfnamefont {G.}~\bibnamefont {Ribordy}},
  \bibinfo {author} {\bibfnamefont {W.}~\bibnamefont {Tittel}},\ and\ \bibinfo
  {author} {\bibfnamefont {H.}~\bibnamefont {Zbinden}},\ }\bibinfo {title}
  {Quantum cryptography},\ \href@noop {} {\bibfield  {journal} {\bibinfo
  {journal} {Rev. Mod. Phys.}\ }\textbf {\bibinfo {volume} {74}},\ \bibinfo
  {pages} {145} (\bibinfo {year} {2002})}\BibitemShut {NoStop}%
\bibitem [{\citenamefont {Pirandola}\ \emph {et~al.}(2020)\citenamefont
  {Pirandola}, \citenamefont {Andersen}, \citenamefont {Banchi}, \citenamefont
  {Berta}, \citenamefont {Bunandar}, \citenamefont {Colbeck}, \citenamefont
  {Englund}, \citenamefont {Gehring}, \citenamefont {Lupo}, \citenamefont
  {Ottaviani} \emph {et~al.}}]{R4_QKD4}%
  \BibitemOpen
  \bibfield  {author} {\bibinfo {author} {\bibfnamefont {S.}~\bibnamefont
  {Pirandola}}, \bibinfo {author} {\bibfnamefont {U.~L.}\ \bibnamefont
  {Andersen}}, \bibinfo {author} {\bibfnamefont {L.}~\bibnamefont {Banchi}},
  \bibinfo {author} {\bibfnamefont {M.}~\bibnamefont {Berta}}, \bibinfo
  {author} {\bibfnamefont {D.}~\bibnamefont {Bunandar}}, \bibinfo {author}
  {\bibfnamefont {R.}~\bibnamefont {Colbeck}}, \bibinfo {author} {\bibfnamefont
  {D.}~\bibnamefont {Englund}}, \bibinfo {author} {\bibfnamefont
  {T.}~\bibnamefont {Gehring}}, \bibinfo {author} {\bibfnamefont
  {C.}~\bibnamefont {Lupo}}, \bibinfo {author} {\bibfnamefont {C.}~\bibnamefont
  {Ottaviani}}, \emph {et~al.},\ }\bibinfo {title} {Advances in quantum
  cryptography},\ \href@noop {} {\bibfield  {journal} {\bibinfo  {journal}
  {Adv. Opt. Photonics}\ }\textbf {\bibinfo {volume} {12}},\ \bibinfo {pages}
  {1012} (\bibinfo {year} {2020})}\BibitemShut {NoStop}%
\bibitem [{\citenamefont {Xu}\ \emph {et~al.}(2020)\citenamefont {Xu},
  \citenamefont {Ma}, \citenamefont {Zhang}, \citenamefont {Lo},\ and\
  \citenamefont {Pan}}]{R5_QKD5}%
  \BibitemOpen
  \bibfield  {author} {\bibinfo {author} {\bibfnamefont {F.}~\bibnamefont
  {Xu}}, \bibinfo {author} {\bibfnamefont {X.}~\bibnamefont {Ma}}, \bibinfo
  {author} {\bibfnamefont {Q.}~\bibnamefont {Zhang}}, \bibinfo {author}
  {\bibfnamefont {H.-K.}\ \bibnamefont {Lo}},\ and\ \bibinfo {author}
  {\bibfnamefont {J.-W.}\ \bibnamefont {Pan}},\ }\bibinfo {title} {Secure
  quantum key distribution with realistic devices},\ \href@noop {} {\bibfield
  {journal} {\bibinfo  {journal} {Rev. Mod. Phys.}\ }\textbf {\bibinfo {volume}
  {92}},\ \bibinfo {pages} {025002} (\bibinfo {year} {2020})}\BibitemShut
  {NoStop}%
\bibitem [{\citenamefont {Shannon}(1949)}]{R6_InformationSecurity}%
  \BibitemOpen
  \bibfield  {author} {\bibinfo {author} {\bibfnamefont {C.~E.}\ \bibnamefont
  {Shannon}},\ }\bibinfo {title} {Communication theory of secrecy systems},\
  \href@noop {} {\bibfield  {journal} {\bibinfo  {journal} {Bell Syst. Tech.
  J.}\ }\textbf {\bibinfo {volume} {28}},\ \bibinfo {pages} {656} (\bibinfo
  {year} {1949})}\BibitemShut {NoStop}%
\bibitem [{\citenamefont {Shor}(1994)}]{R7_Shor}%
  \BibitemOpen
  \bibfield  {author} {\bibinfo {author} {\bibfnamefont {P.~W.}\ \bibnamefont
  {Shor}},\ }\bibinfo {title} {Algorithms for quantum computation: discrete
  logarithms and factoring},\ in\ \href@noop {} {\emph {\bibinfo {booktitle}
  {Proceedings 35th annual symposium on foundations of computer science}}}\
  (\bibinfo {organization} {Ieee},\ \bibinfo {year} {1994})\ pp.\ \bibinfo
  {pages} {124--134}\BibitemShut {NoStop}%
\bibitem [{\citenamefont {Grosshans}\ and\ \citenamefont
  {Grangier}(2002)}]{R8_CVQKD1}%
  \BibitemOpen
  \bibfield  {author} {\bibinfo {author} {\bibfnamefont {F.}~\bibnamefont
  {Grosshans}}\ and\ \bibinfo {author} {\bibfnamefont {P.}~\bibnamefont
  {Grangier}},\ }\bibinfo {title} {Continuous variable quantum cryptography
  using coherent states},\ \href@noop {} {\bibfield  {journal} {\bibinfo
  {journal} {Phys. Rev. Lett.}\ }\textbf {\bibinfo {volume} {88}},\ \bibinfo
  {pages} {057902} (\bibinfo {year} {2002})}\BibitemShut {NoStop}%
\bibitem [{\citenamefont {Weedbrook}\ \emph {et~al.}(2004)\citenamefont
  {Weedbrook}, \citenamefont {Lance}, \citenamefont {Bowen}, \citenamefont
  {Symul}, \citenamefont {Ralph},\ and\ \citenamefont {Lam}}]{R9_CVQKD2}%
  \BibitemOpen
  \bibfield  {author} {\bibinfo {author} {\bibfnamefont {C.}~\bibnamefont
  {Weedbrook}}, \bibinfo {author} {\bibfnamefont {A.~M.}\ \bibnamefont
  {Lance}}, \bibinfo {author} {\bibfnamefont {W.~P.}\ \bibnamefont {Bowen}},
  \bibinfo {author} {\bibfnamefont {T.}~\bibnamefont {Symul}}, \bibinfo
  {author} {\bibfnamefont {T.~C.}\ \bibnamefont {Ralph}},\ and\ \bibinfo
  {author} {\bibfnamefont {P.~K.}\ \bibnamefont {Lam}},\ }\bibinfo {title}
  {Quantum cryptography without switching},\ \href@noop {} {\bibfield
  {journal} {\bibinfo  {journal} {Phys. Rev. Lett.}\ }\textbf {\bibinfo
  {volume} {93}},\ \bibinfo {pages} {170504} (\bibinfo {year}
  {2004})}\BibitemShut {NoStop}%
\bibitem [{\citenamefont {Leverrier}\ \emph {et~al.}(2013)\citenamefont
  {Leverrier}, \citenamefont {Garc{\'\i}a-Patr{\'o}n}, \citenamefont {Renner},\
  and\ \citenamefont {Cerf}}]{R10_CVQKD_T1}%
  \BibitemOpen
  \bibfield  {author} {\bibinfo {author} {\bibfnamefont {A.}~\bibnamefont
  {Leverrier}}, \bibinfo {author} {\bibfnamefont {R.}~\bibnamefont
  {Garc{\'\i}a-Patr{\'o}n}}, \bibinfo {author} {\bibfnamefont {R.}~\bibnamefont
  {Renner}},\ and\ \bibinfo {author} {\bibfnamefont {N.~J.}\ \bibnamefont
  {Cerf}},\ }\bibinfo {title} {Security of continuous-variable quantum key
  distribution against general attacks},\ \href@noop {} {\bibfield  {journal}
  {\bibinfo  {journal} {Phys. Rev. Lett.}\ }\textbf {\bibinfo {volume} {110}},\
  \bibinfo {pages} {030502} (\bibinfo {year} {2013})}\BibitemShut {NoStop}%
\bibitem [{\citenamefont {Leverrier}(2015)}]{R11_CVQKD_T2}%
  \BibitemOpen
  \bibfield  {author} {\bibinfo {author} {\bibfnamefont {A.}~\bibnamefont
  {Leverrier}},\ }\bibinfo {title} {Composable security proof for
  continuous-variable quantum key distribution with coherent states},\
  \href@noop {} {\bibfield  {journal} {\bibinfo  {journal} {Phys. Rev. Lett.}\
  }\textbf {\bibinfo {volume} {114}},\ \bibinfo {pages} {070501} (\bibinfo
  {year} {2015})}\BibitemShut {NoStop}%
\bibitem [{\citenamefont {Leverrier}(2017)}]{R12_CVQKD_T3}%
  \BibitemOpen
  \bibfield  {author} {\bibinfo {author} {\bibfnamefont {A.}~\bibnamefont
  {Leverrier}},\ }\bibinfo {title} {Security of continuous-variable quantum key
  distribution via a gaussian de finetti reduction},\ \href@noop {} {\bibfield
  {journal} {\bibinfo  {journal} {Phys. Rev. Lett.}\ }\textbf {\bibinfo
  {volume} {118}},\ \bibinfo {pages} {200501} (\bibinfo {year}
  {2017})}\BibitemShut {NoStop}%
\bibitem [{\citenamefont {Pirandola}\ \emph {et~al.}(2017)\citenamefont
  {Pirandola}, \citenamefont {Laurenza}, \citenamefont {Ottaviani},\ and\
  \citenamefont {Banchi}}]{R13_CVQKD_T4}%
  \BibitemOpen
  \bibfield  {author} {\bibinfo {author} {\bibfnamefont {S.}~\bibnamefont
  {Pirandola}}, \bibinfo {author} {\bibfnamefont {R.}~\bibnamefont {Laurenza}},
  \bibinfo {author} {\bibfnamefont {C.}~\bibnamefont {Ottaviani}},\ and\
  \bibinfo {author} {\bibfnamefont {L.}~\bibnamefont {Banchi}},\ }\bibinfo
  {title} {Fundamental limits of repeaterless quantum communications},\
  \href@noop {} {\bibfield  {journal} {\bibinfo  {journal} {Nat. Commun.}\
  }\textbf {\bibinfo {volume} {8}},\ \bibinfo {pages} {1} (\bibinfo {year}
  {2017})}\BibitemShut {NoStop}%
\bibitem [{\citenamefont {Lupo}\ \emph {et~al.}(2018)\citenamefont {Lupo},
  \citenamefont {Ottaviani}, \citenamefont {Papanastasiou},\ and\ \citenamefont
  {Pirandola}}]{R14_CVQKD_T5}%
  \BibitemOpen
  \bibfield  {author} {\bibinfo {author} {\bibfnamefont {C.}~\bibnamefont
  {Lupo}}, \bibinfo {author} {\bibfnamefont {C.}~\bibnamefont {Ottaviani}},
  \bibinfo {author} {\bibfnamefont {P.}~\bibnamefont {Papanastasiou}},\ and\
  \bibinfo {author} {\bibfnamefont {S.}~\bibnamefont {Pirandola}},\ }\bibinfo
  {title} {Parameter estimation with almost no public communication for
  continuous-variable quantum key distribution},\ \href@noop {} {\bibfield
  {journal} {\bibinfo  {journal} {Phys. Rev. Lett.}\ }\textbf {\bibinfo
  {volume} {120}},\ \bibinfo {pages} {220505} (\bibinfo {year}
  {2018})}\BibitemShut {NoStop}%
\bibitem [{\citenamefont {Ghorai}\ \emph {et~al.}(2019)\citenamefont {Ghorai},
  \citenamefont {Grangier}, \citenamefont {Diamanti},\ and\ \citenamefont
  {Leverrier}}]{R15_CVQKD_T6}%
  \BibitemOpen
  \bibfield  {author} {\bibinfo {author} {\bibfnamefont {S.}~\bibnamefont
  {Ghorai}}, \bibinfo {author} {\bibfnamefont {P.}~\bibnamefont {Grangier}},
  \bibinfo {author} {\bibfnamefont {E.}~\bibnamefont {Diamanti}},\ and\
  \bibinfo {author} {\bibfnamefont {A.}~\bibnamefont {Leverrier}},\ }\bibinfo
  {title} {Asymptotic security of continuous-variable quantum key distribution
  with a discrete modulation},\ \href@noop {} {\bibfield  {journal} {\bibinfo
  {journal} {Phys. Rev. X}\ }\textbf {\bibinfo {volume} {9}},\ \bibinfo {pages}
  {021059} (\bibinfo {year} {2019})}\BibitemShut {NoStop}%
\bibitem [{\citenamefont {Lin}\ \emph {et~al.}(2019)\citenamefont {Lin},
  \citenamefont {Upadhyaya},\ and\ \citenamefont
  {L{\"u}tkenhaus}}]{R16_CVQKD_T7}%
  \BibitemOpen
  \bibfield  {author} {\bibinfo {author} {\bibfnamefont {J.}~\bibnamefont
  {Lin}}, \bibinfo {author} {\bibfnamefont {T.}~\bibnamefont {Upadhyaya}},\
  and\ \bibinfo {author} {\bibfnamefont {N.}~\bibnamefont {L{\"u}tkenhaus}},\
  }\bibinfo {title} {Asymptotic security analysis of discrete-modulated
  continuous-variable quantum key distribution},\ \href@noop {} {\bibfield
  {journal} {\bibinfo  {journal} {Phys. Rev. X}\ }\textbf {\bibinfo {volume}
  {9}},\ \bibinfo {pages} {041064} (\bibinfo {year} {2019})}\BibitemShut
  {NoStop}%
\bibitem [{\citenamefont {Upadhyaya}\ \emph {et~al.}(2021)\citenamefont
  {Upadhyaya}, \citenamefont {van Himbeeck}, \citenamefont {Lin},\ and\
  \citenamefont {L{\"u}tkenhaus}}]{R17_CVQKD_T8}%
  \BibitemOpen
  \bibfield  {author} {\bibinfo {author} {\bibfnamefont {T.}~\bibnamefont
  {Upadhyaya}}, \bibinfo {author} {\bibfnamefont {T.}~\bibnamefont {van
  Himbeeck}}, \bibinfo {author} {\bibfnamefont {J.}~\bibnamefont {Lin}},\ and\
  \bibinfo {author} {\bibfnamefont {N.}~\bibnamefont {L{\"u}tkenhaus}},\
  }\bibinfo {title} {Dimension reduction in quantum key distribution for
  continuous-and discrete-variable protocols},\ \href@noop {} {\bibfield
  {journal} {\bibinfo  {journal} {PRX Quantum}\ }\textbf {\bibinfo {volume}
  {2}},\ \bibinfo {pages} {020325} (\bibinfo {year} {2021})}\BibitemShut
  {NoStop}%
\bibitem [{\citenamefont {Denys}\ \emph {et~al.}(2021)\citenamefont {Denys},
  \citenamefont {Brown},\ and\ \citenamefont {Leverrier}}]{R18_CVQKD_T9}%
  \BibitemOpen
  \bibfield  {author} {\bibinfo {author} {\bibfnamefont {A.}~\bibnamefont
  {Denys}}, \bibinfo {author} {\bibfnamefont {P.}~\bibnamefont {Brown}},\ and\
  \bibinfo {author} {\bibfnamefont {A.}~\bibnamefont {Leverrier}},\ }\bibinfo
  {title} {Explicit asymptotic secret key rate of continuous-variable quantum
  key distribution with an arbitrary modulation},\ \href@noop {} {\bibfield
  {journal} {\bibinfo  {journal} {Quantum}\ }\textbf {\bibinfo {volume} {5}},\
  \bibinfo {pages} {540} (\bibinfo {year} {2021})}\BibitemShut {NoStop}%
\bibitem [{\citenamefont {Lupo}\ and\ \citenamefont
  {Ouyang}(2022)}]{R19_CVQKD_T10}%
  \BibitemOpen
  \bibfield  {author} {\bibinfo {author} {\bibfnamefont {C.}~\bibnamefont
  {Lupo}}\ and\ \bibinfo {author} {\bibfnamefont {Y.}~\bibnamefont {Ouyang}},\
  }\bibinfo {title} {Quantum key distribution with nonideal heterodyne
  detection: composable security of discrete-modulation continuous-variable
  protocols},\ \href@noop {} {\bibfield  {journal} {\bibinfo  {journal} {PRX
  Quantum}\ }\textbf {\bibinfo {volume} {3}},\ \bibinfo {pages} {010341}
  (\bibinfo {year} {2022})}\BibitemShut {NoStop}%
\bibitem [{\citenamefont {Chen}\ \emph {et~al.}(2023)\citenamefont {Chen},
  \citenamefont {Wang}, \citenamefont {Yu}, \citenamefont {Li},\ and\
  \citenamefont {Guo}}]{R20_CVQKD_T11}%
  \BibitemOpen
  \bibfield  {author} {\bibinfo {author} {\bibfnamefont {Z.}~\bibnamefont
  {Chen}}, \bibinfo {author} {\bibfnamefont {X.}~\bibnamefont {Wang}}, \bibinfo
  {author} {\bibfnamefont {S.}~\bibnamefont {Yu}}, \bibinfo {author}
  {\bibfnamefont {Z.}~\bibnamefont {Li}},\ and\ \bibinfo {author}
  {\bibfnamefont {H.}~\bibnamefont {Guo}},\ }\bibinfo {title} {Continuous-mode
  quantum key distribution with digital signal processing},\ \href@noop {}
  {\bibfield  {journal} {\bibinfo  {journal} {npj Quantum Inform.}\ }\textbf
  {\bibinfo {volume} {9}},\ \bibinfo {pages} {28} (\bibinfo {year}
  {2023})}\BibitemShut {NoStop}%
\bibitem [{\citenamefont {Zhang}\ \emph {et~al.}(2023)\citenamefont {Zhang},
  \citenamefont {Liu}, \citenamefont {Qi}, \citenamefont {He},\ and\
  \citenamefont {Huang}}]{R21_CVQKD_T12}%
  \BibitemOpen
  \bibfield  {author} {\bibinfo {author} {\bibfnamefont {Z.-K.}\ \bibnamefont
  {Zhang}}, \bibinfo {author} {\bibfnamefont {W.-Q.}\ \bibnamefont {Liu}},
  \bibinfo {author} {\bibfnamefont {J.}~\bibnamefont {Qi}}, \bibinfo {author}
  {\bibfnamefont {C.}~\bibnamefont {He}},\ and\ \bibinfo {author}
  {\bibfnamefont {P.}~\bibnamefont {Huang}},\ }\bibinfo {title} {Automatic
  phase compensation of a continuous-variable quantum-key-distribution system
  via deep learning},\ \href@noop {} {\bibfield  {journal} {\bibinfo  {journal}
  {Phys. Rev. A}\ }\textbf {\bibinfo {volume} {107}},\ \bibinfo {pages}
  {062614} (\bibinfo {year} {2023})}\BibitemShut {NoStop}%
\bibitem [{\citenamefont {Kanitschar}\ \emph {et~al.}(2023)\citenamefont
  {Kanitschar}, \citenamefont {George}, \citenamefont {Lin}, \citenamefont
  {Upadhyaya},\ and\ \citenamefont {L{\"u}tkenhaus}}]{R22_CVQKD_T13}%
  \BibitemOpen
  \bibfield  {author} {\bibinfo {author} {\bibfnamefont {F.}~\bibnamefont
  {Kanitschar}}, \bibinfo {author} {\bibfnamefont {I.}~\bibnamefont {George}},
  \bibinfo {author} {\bibfnamefont {J.}~\bibnamefont {Lin}}, \bibinfo {author}
  {\bibfnamefont {T.}~\bibnamefont {Upadhyaya}},\ and\ \bibinfo {author}
  {\bibfnamefont {N.}~\bibnamefont {L{\"u}tkenhaus}},\ }\bibinfo {title}
  {Finite-size security for discrete-modulated continuous-variable quantum key
  distribution protocols},\ \href@noop {} {\bibfield  {journal} {\bibinfo
  {journal} {PRX Quantum}\ }\textbf {\bibinfo {volume} {4}},\ \bibinfo {pages}
  {040306} (\bibinfo {year} {2023})}\BibitemShut {NoStop}%
\bibitem [{\citenamefont {Jouguet}\ \emph {et~al.}(2013)\citenamefont
  {Jouguet}, \citenamefont {Kunz-Jacques}, \citenamefont {Leverrier},
  \citenamefont {Grangier},\ and\ \citenamefont {Diamanti}}]{R23_CVQKD_E1}%
  \BibitemOpen
  \bibfield  {author} {\bibinfo {author} {\bibfnamefont {P.}~\bibnamefont
  {Jouguet}}, \bibinfo {author} {\bibfnamefont {S.}~\bibnamefont
  {Kunz-Jacques}}, \bibinfo {author} {\bibfnamefont {A.}~\bibnamefont
  {Leverrier}}, \bibinfo {author} {\bibfnamefont {P.}~\bibnamefont
  {Grangier}},\ and\ \bibinfo {author} {\bibfnamefont {E.}~\bibnamefont
  {Diamanti}},\ }\bibinfo {title} {Experimental demonstration of long-distance
  continuous-variable quantum key distribution},\ \href@noop {} {\bibfield
  {journal} {\bibinfo  {journal} {Nat. Photonics}\ }\textbf {\bibinfo {volume}
  {7}},\ \bibinfo {pages} {378} (\bibinfo {year} {2013})}\BibitemShut {NoStop}%
\bibitem [{\citenamefont {Qi}\ \emph {et~al.}(2015)\citenamefont {Qi},
  \citenamefont {Lougovski}, \citenamefont {Pooser}, \citenamefont {Grice},\
  and\ \citenamefont {Bobrek}}]{R24_CVQKD_E2}%
  \BibitemOpen
  \bibfield  {author} {\bibinfo {author} {\bibfnamefont {B.}~\bibnamefont
  {Qi}}, \bibinfo {author} {\bibfnamefont {P.}~\bibnamefont {Lougovski}},
  \bibinfo {author} {\bibfnamefont {R.}~\bibnamefont {Pooser}}, \bibinfo
  {author} {\bibfnamefont {W.}~\bibnamefont {Grice}},\ and\ \bibinfo {author}
  {\bibfnamefont {M.}~\bibnamefont {Bobrek}},\ }\bibinfo {title} {Generating
  the local oscillator “locally” in continuous-variable quantum key
  distribution based on coherent detection},\ \href@noop {} {\bibfield
  {journal} {\bibinfo  {journal} {Phys. Rev. X}\ }\textbf {\bibinfo {volume}
  {5}},\ \bibinfo {pages} {041009} (\bibinfo {year} {2015})}\BibitemShut
  {NoStop}%
\bibitem [{\citenamefont {Soh}\ \emph {et~al.}(2015)\citenamefont {Soh},
  \citenamefont {Brif}, \citenamefont {Coles}, \citenamefont {L{\"u}tkenhaus},
  \citenamefont {Camacho}, \citenamefont {Urayama},\ and\ \citenamefont
  {Sarovar}}]{R25_CVQKD_E3}%
  \BibitemOpen
  \bibfield  {author} {\bibinfo {author} {\bibfnamefont {D.~B.}\ \bibnamefont
  {Soh}}, \bibinfo {author} {\bibfnamefont {C.}~\bibnamefont {Brif}}, \bibinfo
  {author} {\bibfnamefont {P.~J.}\ \bibnamefont {Coles}}, \bibinfo {author}
  {\bibfnamefont {N.}~\bibnamefont {L{\"u}tkenhaus}}, \bibinfo {author}
  {\bibfnamefont {R.~M.}\ \bibnamefont {Camacho}}, \bibinfo {author}
  {\bibfnamefont {J.}~\bibnamefont {Urayama}},\ and\ \bibinfo {author}
  {\bibfnamefont {M.}~\bibnamefont {Sarovar}},\ }\bibinfo {title}
  {Self-referenced continuous-variable quantum key distribution protocol},\
  \href@noop {} {\bibfield  {journal} {\bibinfo  {journal} {Phys. Rev. X}\
  }\textbf {\bibinfo {volume} {5}},\ \bibinfo {pages} {041010} (\bibinfo {year}
  {2015})}\BibitemShut {NoStop}%
\bibitem [{\citenamefont {Huang}\ \emph {et~al.}(2016)\citenamefont {Huang},
  \citenamefont {Huang}, \citenamefont {Lin},\ and\ \citenamefont
  {Zeng}}]{R26_CVQKD_E4}%
  \BibitemOpen
  \bibfield  {author} {\bibinfo {author} {\bibfnamefont {D.}~\bibnamefont
  {Huang}}, \bibinfo {author} {\bibfnamefont {P.}~\bibnamefont {Huang}},
  \bibinfo {author} {\bibfnamefont {D.}~\bibnamefont {Lin}},\ and\ \bibinfo
  {author} {\bibfnamefont {G.}~\bibnamefont {Zeng}},\ }\bibinfo {title}
  {Long-distance continuous-variable quantum key distribution by controlling
  excess noise},\ \href@noop {} {\bibfield  {journal} {\bibinfo  {journal}
  {Sci. Rep.}\ }\textbf {\bibinfo {volume} {6}},\ \bibinfo {pages} {19201}
  (\bibinfo {year} {2016})}\BibitemShut {NoStop}%
\bibitem [{\citenamefont {Ren}\ \emph {et~al.}(2021)\citenamefont {Ren},
  \citenamefont {Yang}, \citenamefont {Wonfor}, \citenamefont {White},\ and\
  \citenamefont {Penty}}]{R27_CVQKD_E5}%
  \BibitemOpen
  \bibfield  {author} {\bibinfo {author} {\bibfnamefont {S.}~\bibnamefont
  {Ren}}, \bibinfo {author} {\bibfnamefont {S.}~\bibnamefont {Yang}}, \bibinfo
  {author} {\bibfnamefont {A.}~\bibnamefont {Wonfor}}, \bibinfo {author}
  {\bibfnamefont {I.}~\bibnamefont {White}},\ and\ \bibinfo {author}
  {\bibfnamefont {R.}~\bibnamefont {Penty}},\ }\bibinfo {title} {Demonstration
  of high-speed and low-complexity continuous variable quantum key distribution
  system with local local oscillator},\ \href@noop {} {\bibfield  {journal}
  {\bibinfo  {journal} {Scientific Reports}\ }\textbf {\bibinfo {volume}
  {11}},\ \bibinfo {pages} {9454} (\bibinfo {year} {2021})}\BibitemShut
  {NoStop}%
\bibitem [{\citenamefont {Tian}\ \emph {et~al.}(2022)\citenamefont {Tian},
  \citenamefont {Wang}, \citenamefont {Liu}, \citenamefont {Du}, \citenamefont
  {Liu}, \citenamefont {Lu}, \citenamefont {Wang},\ and\ \citenamefont
  {Li}}]{R28_CVQKD_E6}%
  \BibitemOpen
  \bibfield  {author} {\bibinfo {author} {\bibfnamefont {Y.}~\bibnamefont
  {Tian}}, \bibinfo {author} {\bibfnamefont {P.}~\bibnamefont {Wang}}, \bibinfo
  {author} {\bibfnamefont {J.}~\bibnamefont {Liu}}, \bibinfo {author}
  {\bibfnamefont {S.}~\bibnamefont {Du}}, \bibinfo {author} {\bibfnamefont
  {W.}~\bibnamefont {Liu}}, \bibinfo {author} {\bibfnamefont {Z.}~\bibnamefont
  {Lu}}, \bibinfo {author} {\bibfnamefont {X.}~\bibnamefont {Wang}},\ and\
  \bibinfo {author} {\bibfnamefont {Y.}~\bibnamefont {Li}},\ }\bibinfo {title}
  {Experimental demonstration of continuous-variable
  measurement-device-independent quantum key distribution over optical fiber},\
  \href@noop {} {\bibfield  {journal} {\bibinfo  {journal} {Optica}\ }\textbf
  {\bibinfo {volume} {9}},\ \bibinfo {pages} {492} (\bibinfo {year}
  {2022})}\BibitemShut {NoStop}%
\bibitem [{\citenamefont {Pan}\ \emph {et~al.}(2022)\citenamefont {Pan},
  \citenamefont {Wang}, \citenamefont {Shao}, \citenamefont {Pi}, \citenamefont
  {Li}, \citenamefont {Liu}, \citenamefont {Huang},\ and\ \citenamefont
  {Xu}}]{R29_CVQKD_E7}%
  \BibitemOpen
  \bibfield  {author} {\bibinfo {author} {\bibfnamefont {Y.}~\bibnamefont
  {Pan}}, \bibinfo {author} {\bibfnamefont {H.}~\bibnamefont {Wang}}, \bibinfo
  {author} {\bibfnamefont {Y.}~\bibnamefont {Shao}}, \bibinfo {author}
  {\bibfnamefont {Y.}~\bibnamefont {Pi}}, \bibinfo {author} {\bibfnamefont
  {Y.}~\bibnamefont {Li}}, \bibinfo {author} {\bibfnamefont {B.}~\bibnamefont
  {Liu}}, \bibinfo {author} {\bibfnamefont {W.}~\bibnamefont {Huang}},\ and\
  \bibinfo {author} {\bibfnamefont {B.}~\bibnamefont {Xu}},\ }\bibinfo {title}
  {Experimental demonstration of high-rate discrete-modulated
  continuous-variable quantum key distribution system},\ \href@noop {}
  {\bibfield  {journal} {\bibinfo  {journal} {Optics Letters}\ }\textbf
  {\bibinfo {volume} {47}},\ \bibinfo {pages} {3307} (\bibinfo {year}
  {2022})}\BibitemShut {NoStop}%
\bibitem [{\citenamefont {Tian}\ \emph {et~al.}(2023)\citenamefont {Tian},
  \citenamefont {Zhang}, \citenamefont {Liu}, \citenamefont {Wang},
  \citenamefont {Lu}, \citenamefont {Wang},\ and\ \citenamefont
  {Li}}]{R30_CVQKD_E8}%
  \BibitemOpen
  \bibfield  {author} {\bibinfo {author} {\bibfnamefont {Y.}~\bibnamefont
  {Tian}}, \bibinfo {author} {\bibfnamefont {Y.}~\bibnamefont {Zhang}},
  \bibinfo {author} {\bibfnamefont {S.}~\bibnamefont {Liu}}, \bibinfo {author}
  {\bibfnamefont {P.}~\bibnamefont {Wang}}, \bibinfo {author} {\bibfnamefont
  {Z.}~\bibnamefont {Lu}}, \bibinfo {author} {\bibfnamefont {X.}~\bibnamefont
  {Wang}},\ and\ \bibinfo {author} {\bibfnamefont {Y.}~\bibnamefont {Li}},\
  }\bibinfo {title} {High-performance long-distance discrete-modulation
  continuous-variable quantum key distribution},\ \href@noop {} {\bibfield
  {journal} {\bibinfo  {journal} {Opt. Lett.}\ }\textbf {\bibinfo {volume}
  {48}},\ \bibinfo {pages} {2953} (\bibinfo {year} {2023})}\BibitemShut
  {NoStop}%
\bibitem [{\citenamefont {Xu}\ \emph {et~al.}(2023{\natexlab{a}})\citenamefont
  {Xu}, \citenamefont {Wang}, \citenamefont {Li}, \citenamefont {Zhao},
  \citenamefont {Huang},\ and\ \citenamefont {Zeng}}]{R31_CVQKD_E9}%
  \BibitemOpen
  \bibfield  {author} {\bibinfo {author} {\bibfnamefont {Y.}~\bibnamefont
  {Xu}}, \bibinfo {author} {\bibfnamefont {T.}~\bibnamefont {Wang}}, \bibinfo
  {author} {\bibfnamefont {L.}~\bibnamefont {Li}}, \bibinfo {author}
  {\bibfnamefont {H.}~\bibnamefont {Zhao}}, \bibinfo {author} {\bibfnamefont
  {P.}~\bibnamefont {Huang}},\ and\ \bibinfo {author} {\bibfnamefont
  {G.}~\bibnamefont {Zeng}},\ }\bibinfo {title} {Simultaneous
  continuous-variable quantum key distribution and classical optical
  communication over a shared infrastructure},\ \href@noop {} {\bibfield
  {journal} {\bibinfo  {journal} {Appl. Phys. Lett.}\ }\textbf {\bibinfo
  {volume} {123}} (\bibinfo {year} {2023}{\natexlab{a}})}\BibitemShut {NoStop}%
\bibitem [{\citenamefont {Williams}\ \emph {et~al.}(2024)\citenamefont
  {Williams}, \citenamefont {Qi}, \citenamefont {Alshowkan}, \citenamefont
  {Evans},\ and\ \citenamefont {Peters}}]{R32_CVQKD_E10}%
  \BibitemOpen
  \bibfield  {author} {\bibinfo {author} {\bibfnamefont {B.~P.}\ \bibnamefont
  {Williams}}, \bibinfo {author} {\bibfnamefont {B.}~\bibnamefont {Qi}},
  \bibinfo {author} {\bibfnamefont {M.}~\bibnamefont {Alshowkan}}, \bibinfo
  {author} {\bibfnamefont {P.~G.}\ \bibnamefont {Evans}},\ and\ \bibinfo
  {author} {\bibfnamefont {N.~A.}\ \bibnamefont {Peters}},\ }\bibinfo {title}
  {Field test of continuous-variable quantum key distribution with a true local
  oscillator},\ \href@noop {} {\bibfield  {journal} {\bibinfo  {journal} {Phys.
  Rev. Appl.}\ }\textbf {\bibinfo {volume} {21}},\ \bibinfo {pages} {014056}
  (\bibinfo {year} {2024})}\BibitemShut {NoStop}%
\bibitem [{\citenamefont {Wang}\ \emph {et~al.}(2024)\citenamefont {Wang},
  \citenamefont {Huang}, \citenamefont {Li}, \citenamefont {Zhou},\ and\
  \citenamefont {Zeng}}]{R33_CVQKD_E11}%
  \BibitemOpen
  \bibfield  {author} {\bibinfo {author} {\bibfnamefont {T.}~\bibnamefont
  {Wang}}, \bibinfo {author} {\bibfnamefont {P.}~\bibnamefont {Huang}},
  \bibinfo {author} {\bibfnamefont {L.}~\bibnamefont {Li}}, \bibinfo {author}
  {\bibfnamefont {Y.}~\bibnamefont {Zhou}},\ and\ \bibinfo {author}
  {\bibfnamefont {G.}~\bibnamefont {Zeng}},\ }\bibinfo {title} {High key rate
  continuous-variable quantum key distribution using telecom optical
  components},\ \href@noop {} {\bibfield  {journal} {\bibinfo  {journal} {New
  Journal of Physics}\ }\textbf {\bibinfo {volume} {26}},\ \bibinfo {pages}
  {023002} (\bibinfo {year} {2024})}\BibitemShut {NoStop}%
\bibitem [{\citenamefont {Xu}\ \emph {et~al.}(2024)\citenamefont {Xu},
  \citenamefont {Wang}, \citenamefont {Liao}, \citenamefont {Zhou},
  \citenamefont {Huang},\ and\ \citenamefont {Zeng}}]{R34_CVQKD_E12}%
  \BibitemOpen
  \bibfield  {author} {\bibinfo {author} {\bibfnamefont {Y.}~\bibnamefont
  {Xu}}, \bibinfo {author} {\bibfnamefont {T.}~\bibnamefont {Wang}}, \bibinfo
  {author} {\bibfnamefont {X.}~\bibnamefont {Liao}}, \bibinfo {author}
  {\bibfnamefont {Y.}~\bibnamefont {Zhou}}, \bibinfo {author} {\bibfnamefont
  {P.}~\bibnamefont {Huang}},\ and\ \bibinfo {author} {\bibfnamefont
  {G.}~\bibnamefont {Zeng}},\ }\bibinfo {title} {Robust continuous-variable
  quantum key distribution in the finite-size regime},\ \href@noop {}
  {\bibfield  {journal} {\bibinfo  {journal} {Photonics Research}\ }\textbf
  {\bibinfo {volume} {12}},\ \bibinfo {pages} {2549} (\bibinfo {year}
  {2024})}\BibitemShut {NoStop}%
\bibitem [{\citenamefont {Liao}\ \emph {et~al.}(2025)\citenamefont {Liao},
  \citenamefont {Xu}, \citenamefont {Zhang}, \citenamefont {Huang},
  \citenamefont {Wang}, \citenamefont {Wang},\ and\ \citenamefont
  {Zeng}}]{R35_CVQKD_E13}%
  \BibitemOpen
  \bibfield  {author} {\bibinfo {author} {\bibfnamefont {X.}~\bibnamefont
  {Liao}}, \bibinfo {author} {\bibfnamefont {Y.}~\bibnamefont {Xu}}, \bibinfo
  {author} {\bibfnamefont {Q.}~\bibnamefont {Zhang}}, \bibinfo {author}
  {\bibfnamefont {P.}~\bibnamefont {Huang}}, \bibinfo {author} {\bibfnamefont
  {T.}~\bibnamefont {Wang}}, \bibinfo {author} {\bibfnamefont {K.}~\bibnamefont
  {Wang}},\ and\ \bibinfo {author} {\bibfnamefont {G.}~\bibnamefont {Zeng}},\
  }\bibinfo {title} {High-rate self-referenced continuous-variable quantum key
  distribution over a high-loss free-space channel},\ \href@noop {} {\bibfield
  {journal} {\bibinfo  {journal} {Photonics Research}\ }\textbf {\bibinfo
  {volume} {13}},\ \bibinfo {pages} {2603} (\bibinfo {year}
  {2025})}\BibitemShut {NoStop}%
\bibitem [{\citenamefont {Ma}\ \emph {et~al.}(2025)\citenamefont {Ma},
  \citenamefont {Qi}, \citenamefont {Cui}, \citenamefont {Shen}, \citenamefont
  {Chen}, \citenamefont {Sun}, \citenamefont {Yu},\ and\ \citenamefont
  {Wang}}]{R36_CVQKD_E14}%
  \BibitemOpen
  \bibfield  {author} {\bibinfo {author} {\bibfnamefont {J.}~\bibnamefont
  {Ma}}, \bibinfo {author} {\bibfnamefont {D.}~\bibnamefont {Qi}}, \bibinfo
  {author} {\bibfnamefont {L.}~\bibnamefont {Cui}}, \bibinfo {author}
  {\bibfnamefont {T.}~\bibnamefont {Shen}}, \bibinfo {author} {\bibfnamefont
  {Z.}~\bibnamefont {Chen}}, \bibinfo {author} {\bibfnamefont {Y.}~\bibnamefont
  {Sun}}, \bibinfo {author} {\bibfnamefont {S.}~\bibnamefont {Yu}},\ and\
  \bibinfo {author} {\bibfnamefont {X.}~\bibnamefont {Wang}},\ }\bibinfo
  {title} {Enhanced-rate continuous-variable quantum key distribution with
  particle filter-based carrier phase recovery},\ \href@noop {} {\bibfield
  {journal} {\bibinfo  {journal} {Optics Express}\ }\textbf {\bibinfo {volume}
  {33}},\ \bibinfo {pages} {47178} (\bibinfo {year} {2025})}\BibitemShut
  {NoStop}%
\bibitem [{\citenamefont {Wang}\ \emph {et~al.}(2022)\citenamefont {Wang},
  \citenamefont {Wang}, \citenamefont {Zhou}, \citenamefont {Chen},
  \citenamefont {Yu},\ and\ \citenamefont {Guo}}]{R37_CVQKD_P1}%
  \BibitemOpen
  \bibfield  {author} {\bibinfo {author} {\bibfnamefont {X.}~\bibnamefont
  {Wang}}, \bibinfo {author} {\bibfnamefont {H.}~\bibnamefont {Wang}}, \bibinfo
  {author} {\bibfnamefont {C.}~\bibnamefont {Zhou}}, \bibinfo {author}
  {\bibfnamefont {Z.}~\bibnamefont {Chen}}, \bibinfo {author} {\bibfnamefont
  {S.}~\bibnamefont {Yu}},\ and\ \bibinfo {author} {\bibfnamefont
  {H.}~\bibnamefont {Guo}},\ }\bibinfo {title} {Continuous-variable quantum key
  distribution with low-complexity information reconciliation},\ \href@noop {}
  {\bibfield  {journal} {\bibinfo  {journal} {Opt. Express}\ }\textbf {\bibinfo
  {volume} {30}},\ \bibinfo {pages} {30455} (\bibinfo {year}
  {2022})}\BibitemShut {NoStop}%
\bibitem [{\citenamefont {Cao}\ \emph {et~al.}(2023)\citenamefont {Cao},
  \citenamefont {Chen}, \citenamefont {Chai}, \citenamefont {Liang},\ and\
  \citenamefont {Yuan}}]{R38_CVQKD_P2}%
  \BibitemOpen
  \bibfield  {author} {\bibinfo {author} {\bibfnamefont {Z.}~\bibnamefont
  {Cao}}, \bibinfo {author} {\bibfnamefont {X.}~\bibnamefont {Chen}}, \bibinfo
  {author} {\bibfnamefont {G.}~\bibnamefont {Chai}}, \bibinfo {author}
  {\bibfnamefont {K.}~\bibnamefont {Liang}},\ and\ \bibinfo {author}
  {\bibfnamefont {Y.}~\bibnamefont {Yuan}},\ }\bibinfo {title} {Rate-adaptive
  polar-coding-based reconciliation for continuous-variable quantum key
  distribution at low signal-to-noise ratio},\ \href@noop {} {\bibfield
  {journal} {\bibinfo  {journal} {Phys. Rev. Appl.}\ }\textbf {\bibinfo
  {volume} {19}},\ \bibinfo {pages} {044023} (\bibinfo {year}
  {2023})}\BibitemShut {NoStop}%
\bibitem [{\citenamefont {Wang}\ \emph
  {et~al.}(2023{\natexlab{a}})\citenamefont {Wang}, \citenamefont {Xu},
  \citenamefont {Zhao}, \citenamefont {Chen}, \citenamefont {Yu},\ and\
  \citenamefont {Guo}}]{R39_CVQKD_P3}%
  \BibitemOpen
  \bibfield  {author} {\bibinfo {author} {\bibfnamefont {X.}~\bibnamefont
  {Wang}}, \bibinfo {author} {\bibfnamefont {M.}~\bibnamefont {Xu}}, \bibinfo
  {author} {\bibfnamefont {Y.}~\bibnamefont {Zhao}}, \bibinfo {author}
  {\bibfnamefont {Z.}~\bibnamefont {Chen}}, \bibinfo {author} {\bibfnamefont
  {S.}~\bibnamefont {Yu}},\ and\ \bibinfo {author} {\bibfnamefont
  {H.}~\bibnamefont {Guo}},\ }\bibinfo {title} {Non-gaussian reconciliation for
  continuous-variable quantum key distribution},\ \href@noop {} {\bibfield
  {journal} {\bibinfo  {journal} {Phys. Rev. Appl.}\ }\textbf {\bibinfo
  {volume} {19}},\ \bibinfo {pages} {054084} (\bibinfo {year}
  {2023}{\natexlab{a}})}\BibitemShut {NoStop}%
\bibitem [{\citenamefont {Xing}\ \emph
  {et~al.}(2025{\natexlab{a}})\citenamefont {Xing}, \citenamefont {Zhou},
  \citenamefont {Ma}, \citenamefont {Chen}, \citenamefont {Yu},\ and\
  \citenamefont {Wang}}]{R40_CVQKD_P4}%
  \BibitemOpen
  \bibfield  {author} {\bibinfo {author} {\bibfnamefont {L.}~\bibnamefont
  {Xing}}, \bibinfo {author} {\bibfnamefont {C.}~\bibnamefont {Zhou}}, \bibinfo
  {author} {\bibfnamefont {J.}~\bibnamefont {Ma}}, \bibinfo {author}
  {\bibfnamefont {Z.}~\bibnamefont {Chen}}, \bibinfo {author} {\bibfnamefont
  {S.}~\bibnamefont {Yu}},\ and\ \bibinfo {author} {\bibfnamefont
  {X.}~\bibnamefont {Wang}},\ }\bibinfo {title} {Information reconciliation
  with extremely low signal-to-noise ratio for continuous-variable quantum key
  distribution with ldpc-hadamard codes},\ \href@noop {} {\bibfield  {journal}
  {\bibinfo  {journal} {Physical Review Applied}\ }\textbf {\bibinfo {volume}
  {24}},\ \bibinfo {pages} {014018} (\bibinfo {year}
  {2025}{\natexlab{a}})}\BibitemShut {NoStop}%
\bibitem [{\citenamefont {Xing}\ \emph
  {et~al.}(2025{\natexlab{b}})\citenamefont {Xing}, \citenamefont {Qi},
  \citenamefont {Chen}, \citenamefont {Yu},\ and\ \citenamefont
  {Wang}}]{R41_CVQKD_P5}%
  \BibitemOpen
  \bibfield  {author} {\bibinfo {author} {\bibfnamefont {L.}~\bibnamefont
  {Xing}}, \bibinfo {author} {\bibfnamefont {D.}~\bibnamefont {Qi}}, \bibinfo
  {author} {\bibfnamefont {Z.}~\bibnamefont {Chen}}, \bibinfo {author}
  {\bibfnamefont {S.}~\bibnamefont {Yu}},\ and\ \bibinfo {author}
  {\bibfnamefont {X.}~\bibnamefont {Wang}},\ }\bibinfo {title} {Rate-adaptive
  non-binary ldpc code-based information reconciliation protocol for
  continuous-variable quantum key distribution},\ \href@noop {} {\bibfield
  {journal} {\bibinfo  {journal} {Advanced Quantum Technologies}\ ,\ \bibinfo
  {pages} {e00389}} (\bibinfo {year} {2025}{\natexlab{b}})}\BibitemShut
  {NoStop}%
\bibitem [{\citenamefont {Wang}\ \emph
  {et~al.}(2023{\natexlab{b}})\citenamefont {Wang}, \citenamefont {Chen},
  \citenamefont {Li}, \citenamefont {Qi}, \citenamefont {Yu},\ and\
  \citenamefont {Guo}}]{R42_CVQKD_N1}%
  \BibitemOpen
  \bibfield  {author} {\bibinfo {author} {\bibfnamefont {X.}~\bibnamefont
  {Wang}}, \bibinfo {author} {\bibfnamefont {Z.}~\bibnamefont {Chen}}, \bibinfo
  {author} {\bibfnamefont {Z.}~\bibnamefont {Li}}, \bibinfo {author}
  {\bibfnamefont {D.}~\bibnamefont {Qi}}, \bibinfo {author} {\bibfnamefont
  {S.}~\bibnamefont {Yu}},\ and\ \bibinfo {author} {\bibfnamefont
  {H.}~\bibnamefont {Guo}},\ }\bibinfo {title} {Experimental upstream
  transmission of continuous variable quantum key distribution access
  network},\ \href@noop {} {\bibfield  {journal} {\bibinfo  {journal} {Opt.
  Lett.}\ }\textbf {\bibinfo {volume} {48}},\ \bibinfo {pages} {3327} (\bibinfo
  {year} {2023}{\natexlab{b}})}\BibitemShut {NoStop}%
\bibitem [{\citenamefont {Xu}\ \emph {et~al.}(2023{\natexlab{b}})\citenamefont
  {Xu}, \citenamefont {Wang}, \citenamefont {Zhao}, \citenamefont {Huang},\
  and\ \citenamefont {Zeng}}]{R43_CVQKD_N2}%
  \BibitemOpen
  \bibfield  {author} {\bibinfo {author} {\bibfnamefont {Y.}~\bibnamefont
  {Xu}}, \bibinfo {author} {\bibfnamefont {T.}~\bibnamefont {Wang}}, \bibinfo
  {author} {\bibfnamefont {H.}~\bibnamefont {Zhao}}, \bibinfo {author}
  {\bibfnamefont {P.}~\bibnamefont {Huang}},\ and\ \bibinfo {author}
  {\bibfnamefont {G.}~\bibnamefont {Zeng}},\ }\bibinfo {title} {Round-trip
  multi-band quantum access network},\ \href@noop {} {\bibfield  {journal}
  {\bibinfo  {journal} {Photonics Res.}\ }\textbf {\bibinfo {volume} {11}},\
  \bibinfo {pages} {1449} (\bibinfo {year} {2023}{\natexlab{b}})}\BibitemShut
  {NoStop}%
\bibitem [{\citenamefont {Qi}\ \emph {et~al.}(2024)\citenamefont {Qi},
  \citenamefont {Wang}, \citenamefont {Li}, \citenamefont {Ma}, \citenamefont
  {Chen}, \citenamefont {Lu},\ and\ \citenamefont {Yu}}]{R44_CVQKD_N3}%
  \BibitemOpen
  \bibfield  {author} {\bibinfo {author} {\bibfnamefont {D.}~\bibnamefont
  {Qi}}, \bibinfo {author} {\bibfnamefont {X.}~\bibnamefont {Wang}}, \bibinfo
  {author} {\bibfnamefont {Z.}~\bibnamefont {Li}}, \bibinfo {author}
  {\bibfnamefont {J.}~\bibnamefont {Ma}}, \bibinfo {author} {\bibfnamefont
  {Z.}~\bibnamefont {Chen}}, \bibinfo {author} {\bibfnamefont {Y.}~\bibnamefont
  {Lu}},\ and\ \bibinfo {author} {\bibfnamefont {S.}~\bibnamefont {Yu}},\
  }\bibinfo {title} {Experimental demonstration of a quantum downstream access
  network in continuous variable quantum key distribution with a local local
  oscillator},\ \href@noop {} {\bibfield  {journal} {\bibinfo  {journal}
  {Photonics Res.}\ }\textbf {\bibinfo {volume} {12}},\ \bibinfo {pages} {1262}
  (\bibinfo {year} {2024})}\BibitemShut {NoStop}%
\bibitem [{\citenamefont {Li}\ \emph {et~al.}(2024)\citenamefont {Li},
  \citenamefont {Wang}, \citenamefont {Qi}, \citenamefont {Chen},\ and\
  \citenamefont {Yu}}]{R45_CVQKD_N4}%
  \BibitemOpen
  \bibfield  {author} {\bibinfo {author} {\bibfnamefont {Z.}~\bibnamefont
  {Li}}, \bibinfo {author} {\bibfnamefont {X.}~\bibnamefont {Wang}}, \bibinfo
  {author} {\bibfnamefont {D.}~\bibnamefont {Qi}}, \bibinfo {author}
  {\bibfnamefont {Z.}~\bibnamefont {Chen}},\ and\ \bibinfo {author}
  {\bibfnamefont {S.}~\bibnamefont {Yu}},\ }\bibinfo {title} {Experimental
  implementation of four-user downstream access network continuous-variable
  quantum key distribution},\ \href@noop {} {\bibfield  {journal} {\bibinfo
  {journal} {J. Lightwave Technol.}\ } (\bibinfo {year} {2024})}\BibitemShut
  {NoStop}%
\bibitem [{\citenamefont {Karinou}\ \emph {et~al.}(2018)\citenamefont
  {Karinou}, \citenamefont {Brunner}, \citenamefont {Fung}, \citenamefont
  {Comandar}, \citenamefont {Bettelli}, \citenamefont {Hillerkuss},
  \citenamefont {Kuschnerov}, \citenamefont {Mikroulis}, \citenamefont {Wang},
  \citenamefont {Xie} \emph {et~al.}}]{R46_DSPstage1}%
  \BibitemOpen
  \bibfield  {author} {\bibinfo {author} {\bibfnamefont {F.}~\bibnamefont
  {Karinou}}, \bibinfo {author} {\bibfnamefont {H.~H.}\ \bibnamefont
  {Brunner}}, \bibinfo {author} {\bibfnamefont {C.-H.~F.}\ \bibnamefont
  {Fung}}, \bibinfo {author} {\bibfnamefont {L.~C.}\ \bibnamefont {Comandar}},
  \bibinfo {author} {\bibfnamefont {S.}~\bibnamefont {Bettelli}}, \bibinfo
  {author} {\bibfnamefont {D.}~\bibnamefont {Hillerkuss}}, \bibinfo {author}
  {\bibfnamefont {M.}~\bibnamefont {Kuschnerov}}, \bibinfo {author}
  {\bibfnamefont {S.}~\bibnamefont {Mikroulis}}, \bibinfo {author}
  {\bibfnamefont {D.}~\bibnamefont {Wang}}, \bibinfo {author} {\bibfnamefont
  {C.}~\bibnamefont {Xie}}, \emph {et~al.},\ }\bibinfo {title} {Toward the
  integration of cv quantum key distribution in deployed optical networks},\
  \href@noop {} {\bibfield  {journal} {\bibinfo  {journal} {IEEE Photonics
  Technology Letters}\ }\textbf {\bibinfo {volume} {30}},\ \bibinfo {pages}
  {650} (\bibinfo {year} {2018})}\BibitemShut {NoStop}%
\bibitem [{\citenamefont {Eriksson}\ \emph {et~al.}(2019)\citenamefont
  {Eriksson}, \citenamefont {Hirano}, \citenamefont {Puttnam}, \citenamefont
  {Rademacher}, \citenamefont {Lu{\'\i}s}, \citenamefont {Fujiwara},
  \citenamefont {Namiki}, \citenamefont {Awaji}, \citenamefont {Takeoka},
  \citenamefont {Wada} \emph {et~al.}}]{R47_DSPstage2}%
  \BibitemOpen
  \bibfield  {author} {\bibinfo {author} {\bibfnamefont {T.~A.}\ \bibnamefont
  {Eriksson}}, \bibinfo {author} {\bibfnamefont {T.}~\bibnamefont {Hirano}},
  \bibinfo {author} {\bibfnamefont {B.~J.}\ \bibnamefont {Puttnam}}, \bibinfo
  {author} {\bibfnamefont {G.}~\bibnamefont {Rademacher}}, \bibinfo {author}
  {\bibfnamefont {R.~S.}\ \bibnamefont {Lu{\'\i}s}}, \bibinfo {author}
  {\bibfnamefont {M.}~\bibnamefont {Fujiwara}}, \bibinfo {author}
  {\bibfnamefont {R.}~\bibnamefont {Namiki}}, \bibinfo {author} {\bibfnamefont
  {Y.}~\bibnamefont {Awaji}}, \bibinfo {author} {\bibfnamefont
  {M.}~\bibnamefont {Takeoka}}, \bibinfo {author} {\bibfnamefont
  {N.}~\bibnamefont {Wada}}, \emph {et~al.},\ }\bibinfo {title} {Wavelength
  division multiplexing of continuous variable quantum key distribution and
  18.3 tbit/s data channels},\ \href@noop {} {\bibfield  {journal} {\bibinfo
  {journal} {Communications Physics}\ }\textbf {\bibinfo {volume} {2}},\
  \bibinfo {pages} {9} (\bibinfo {year} {2019})}\BibitemShut {NoStop}%
\bibitem [{\citenamefont {Eriksson}\ \emph {et~al.}(2020)\citenamefont
  {Eriksson}, \citenamefont {Luis}, \citenamefont {Puttnam}, \citenamefont
  {Rademacher}, \citenamefont {Fujiwara}, \citenamefont {Awaji}, \citenamefont
  {Furukawa}, \citenamefont {Wada}, \citenamefont {Takeoka},\ and\
  \citenamefont {Sasaki}}]{R48_DSPstage3}%
  \BibitemOpen
  \bibfield  {author} {\bibinfo {author} {\bibfnamefont {T.~A.}\ \bibnamefont
  {Eriksson}}, \bibinfo {author} {\bibfnamefont {R.~S.}\ \bibnamefont {Luis}},
  \bibinfo {author} {\bibfnamefont {B.~J.}\ \bibnamefont {Puttnam}}, \bibinfo
  {author} {\bibfnamefont {G.}~\bibnamefont {Rademacher}}, \bibinfo {author}
  {\bibfnamefont {M.}~\bibnamefont {Fujiwara}}, \bibinfo {author}
  {\bibfnamefont {Y.}~\bibnamefont {Awaji}}, \bibinfo {author} {\bibfnamefont
  {H.}~\bibnamefont {Furukawa}}, \bibinfo {author} {\bibfnamefont
  {N.}~\bibnamefont {Wada}}, \bibinfo {author} {\bibfnamefont {M.}~\bibnamefont
  {Takeoka}},\ and\ \bibinfo {author} {\bibfnamefont {M.}~\bibnamefont
  {Sasaki}},\ }\bibinfo {title} {Wavelength division multiplexing of 194
  continuous variable quantum key distribution channels},\ \href@noop {}
  {\bibfield  {journal} {\bibinfo  {journal} {Journal of Lightwave Technology}\
  }\textbf {\bibinfo {volume} {38}},\ \bibinfo {pages} {2214} (\bibinfo {year}
  {2020})}\BibitemShut {NoStop}%
\bibitem [{\citenamefont {Blow}\ \emph {et~al.}(1990)\citenamefont {Blow},
  \citenamefont {Loudon}, \citenamefont {Phoenix},\ and\ \citenamefont
  {Shepherd}}]{R49_CM1}%
  \BibitemOpen
  \bibfield  {author} {\bibinfo {author} {\bibfnamefont {K.}~\bibnamefont
  {Blow}}, \bibinfo {author} {\bibfnamefont {R.}~\bibnamefont {Loudon}},
  \bibinfo {author} {\bibfnamefont {S.~J.}\ \bibnamefont {Phoenix}},\ and\
  \bibinfo {author} {\bibfnamefont {T.}~\bibnamefont {Shepherd}},\ }\bibinfo
  {title} {Continuum fields in quantum optics},\ \href@noop {} {\bibfield
  {journal} {\bibinfo  {journal} {Phys. Rev. A}\ }\textbf {\bibinfo {volume}
  {42}},\ \bibinfo {pages} {4102} (\bibinfo {year} {1990})}\BibitemShut
  {NoStop}%
\bibitem [{\citenamefont {Fabre}\ and\ \citenamefont {Treps}(2020)}]{R50_CM2}%
  \BibitemOpen
  \bibfield  {author} {\bibinfo {author} {\bibfnamefont {C.}~\bibnamefont
  {Fabre}}\ and\ \bibinfo {author} {\bibfnamefont {N.}~\bibnamefont {Treps}},\
  }\bibinfo {title} {Modes and states in quantum optics},\ \href@noop {}
  {\bibfield  {journal} {\bibinfo  {journal} {Rev. Mod. Phys.}\ }\textbf
  {\bibinfo {volume} {92}},\ \bibinfo {pages} {035005} (\bibinfo {year}
  {2020})}\BibitemShut {NoStop}%
\bibitem [{\citenamefont {Raymer}\ \emph {et~al.}(1989)\citenamefont {Raymer},
  \citenamefont {Li},\ and\ \citenamefont {Walmsley}}]{R51_TM1}%
  \BibitemOpen
  \bibfield  {author} {\bibinfo {author} {\bibfnamefont {M.}~\bibnamefont
  {Raymer}}, \bibinfo {author} {\bibfnamefont {Z.}~\bibnamefont {Li}},\ and\
  \bibinfo {author} {\bibfnamefont {I.}~\bibnamefont {Walmsley}},\ }\bibinfo
  {title} {Temporal quantum fluctuations in stimulated raman scattering:
  Coherent-modes description},\ \href@noop {} {\bibfield  {journal} {\bibinfo
  {journal} {Phys. Rev. Lett.}\ }\textbf {\bibinfo {volume} {63}},\ \bibinfo
  {pages} {1586} (\bibinfo {year} {1989})}\BibitemShut {NoStop}%
\bibitem [{\citenamefont {Brecht}\ \emph {et~al.}(2015)\citenamefont {Brecht},
  \citenamefont {Reddy}, \citenamefont {Silberhorn},\ and\ \citenamefont
  {Raymer}}]{R52_TM2}%
  \BibitemOpen
  \bibfield  {author} {\bibinfo {author} {\bibfnamefont {B.}~\bibnamefont
  {Brecht}}, \bibinfo {author} {\bibfnamefont {D.~V.}\ \bibnamefont {Reddy}},
  \bibinfo {author} {\bibfnamefont {C.}~\bibnamefont {Silberhorn}},\ and\
  \bibinfo {author} {\bibfnamefont {M.~G.}\ \bibnamefont {Raymer}},\ }\bibinfo
  {title} {Photon temporal modes: a complete framework for quantum information
  science},\ \href@noop {} {\bibfield  {journal} {\bibinfo  {journal} {Phys.
  Rev. X}\ }\textbf {\bibinfo {volume} {5}},\ \bibinfo {pages} {041017}
  (\bibinfo {year} {2015})}\BibitemShut {NoStop}%
\bibitem [{\citenamefont {Raymer}\ and\ \citenamefont
  {Walmsley}(2020)}]{R53_TM3}%
  \BibitemOpen
  \bibfield  {author} {\bibinfo {author} {\bibfnamefont {M.~G.}\ \bibnamefont
  {Raymer}}\ and\ \bibinfo {author} {\bibfnamefont {I.~A.}\ \bibnamefont
  {Walmsley}},\ }\bibinfo {title} {Temporal modes in quantum optics: then and
  now},\ \href@noop {} {\bibfield  {journal} {\bibinfo  {journal} {Phys. Scr.}\
  }\textbf {\bibinfo {volume} {95}},\ \bibinfo {pages} {064002} (\bibinfo
  {year} {2020})}\BibitemShut {NoStop}%
\bibitem [{\citenamefont {Zhao}\ \emph {et~al.}(2021)\citenamefont {Zhao},
  \citenamefont {Huo}, \citenamefont {Cui}, \citenamefont {Li},\ and\
  \citenamefont {Ou}}]{R54_TM4}%
  \BibitemOpen
  \bibfield  {author} {\bibinfo {author} {\bibfnamefont {W.}~\bibnamefont
  {Zhao}}, \bibinfo {author} {\bibfnamefont {N.}~\bibnamefont {Huo}}, \bibinfo
  {author} {\bibfnamefont {L.}~\bibnamefont {Cui}}, \bibinfo {author}
  {\bibfnamefont {X.}~\bibnamefont {Li}},\ and\ \bibinfo {author}
  {\bibfnamefont {Z.}~\bibnamefont {Ou}},\ }\bibinfo {title} {Propagation of
  temporal mode multiplexed optical fields in fibers: influence of
  dispersion},\ \href@noop {} {\bibfield  {journal} {\bibinfo  {journal} {Opt.
  Express}\ }\textbf {\bibinfo {volume} {30}},\ \bibinfo {pages} {447}
  (\bibinfo {year} {2021})}\BibitemShut {NoStop}%
\bibitem [{\citenamefont {Grosshans}\ \emph {et~al.}(2003)\citenamefont
  {Grosshans}, \citenamefont {Cerf}, \citenamefont {Wenger}, \citenamefont
  {Tualle-Brouri},\ and\ \citenamefont {Grangier}}]{R56_TMSV}%
  \BibitemOpen
  \bibfield  {author} {\bibinfo {author} {\bibfnamefont {F.}~\bibnamefont
  {Grosshans}}, \bibinfo {author} {\bibfnamefont {N.~J.}\ \bibnamefont {Cerf}},
  \bibinfo {author} {\bibfnamefont {J.}~\bibnamefont {Wenger}}, \bibinfo
  {author} {\bibfnamefont {R.}~\bibnamefont {Tualle-Brouri}},\ and\ \bibinfo
  {author} {\bibfnamefont {P.}~\bibnamefont {Grangier}},\ }\bibinfo {title}
  {Virtual entanglement and reconciliation protocols for quantum cryptography
  with continuous variables},\ \href@noop {} {\bibfield  {journal} {\bibinfo
  {journal} {arXiv preprint quant-ph/0306141}\ } (\bibinfo {year}
  {2003})}\BibitemShut {NoStop}%
\bibitem [{\citenamefont {Loudon}(2000)}]{R55_COperator}%
  \BibitemOpen
  \bibfield  {author} {\bibinfo {author} {\bibfnamefont {R.}~\bibnamefont
  {Loudon}},\ }\bibinfo {title} {The quantum theory of light},\ \href@noop {}
  {\emph {\bibinfo {title} {The quantum theory of light}}}\ (\bibinfo
  {publisher} {OUP Oxford},\ \bibinfo {year} {2000})\BibitemShut {NoStop}%
\bibitem [{\citenamefont {Cover}(1999)}]{R57_KeyRate1}%
  \BibitemOpen
  \bibfield  {author} {\bibinfo {author} {\bibfnamefont {T.~M.}\ \bibnamefont
  {Cover}},\ }\bibinfo {title} {Elements of information theory},\ \href@noop {}
  {\emph {\bibinfo {title} {Elements of information theory}}}\ (\bibinfo
  {publisher} {John Wiley \& Sons},\ \bibinfo {year} {1999})\BibitemShut
  {NoStop}%
\bibitem [{\citenamefont {Devetak}\ and\ \citenamefont
  {Winter}(2005)}]{R58_KeyRate2}%
  \BibitemOpen
  \bibfield  {author} {\bibinfo {author} {\bibfnamefont {I.}~\bibnamefont
  {Devetak}}\ and\ \bibinfo {author} {\bibfnamefont {A.}~\bibnamefont
  {Winter}},\ }\bibinfo {title} {Distillation of secret key and entanglement
  from quantum states},\ \href@noop {} {\bibfield  {journal} {\bibinfo
  {journal} {Proceedings of the Royal Society A: Mathematical, Physical and
  engineering sciences}\ }\textbf {\bibinfo {volume} {461}},\ \bibinfo {pages}
  {207} (\bibinfo {year} {2005})}\BibitemShut {NoStop}%
\bibitem [{\citenamefont {Fossier}\ \emph {et~al.}(2009)\citenamefont
  {Fossier}, \citenamefont {Diamanti}, \citenamefont {Debuisschert},
  \citenamefont {Tualle-Brouri},\ and\ \citenamefont {Grangier}}]{R59_DetectM}%
  \BibitemOpen
  \bibfield  {author} {\bibinfo {author} {\bibfnamefont {S.}~\bibnamefont
  {Fossier}}, \bibinfo {author} {\bibfnamefont {E.}~\bibnamefont {Diamanti}},
  \bibinfo {author} {\bibfnamefont {T.}~\bibnamefont {Debuisschert}}, \bibinfo
  {author} {\bibfnamefont {R.}~\bibnamefont {Tualle-Brouri}},\ and\ \bibinfo
  {author} {\bibfnamefont {P.}~\bibnamefont {Grangier}},\ }\bibinfo {title}
  {Improvement of continuous-variable quantum key distribution systems by using
  optical preamplifiers},\ \href@noop {} {\bibfield  {journal} {\bibinfo
  {journal} {Journal of Physics B: Atomic, Molecular and Optical Physics}\
  }\textbf {\bibinfo {volume} {42}},\ \bibinfo {pages} {114014} (\bibinfo
  {year} {2009})}\BibitemShut {NoStop}%
\bibitem [{\citenamefont {Navascu{\'e}s}\ \emph {et~al.}(2006)\citenamefont
  {Navascu{\'e}s}, \citenamefont {Grosshans},\ and\ \citenamefont
  {Ac{\'\i}n}}]{R60_GAttack1}%
  \BibitemOpen
  \bibfield  {author} {\bibinfo {author} {\bibfnamefont {M.}~\bibnamefont
  {Navascu{\'e}s}}, \bibinfo {author} {\bibfnamefont {F.}~\bibnamefont
  {Grosshans}},\ and\ \bibinfo {author} {\bibfnamefont {A.}~\bibnamefont
  {Ac{\'\i}n}},\ }\bibinfo {title} {Optimality of gaussian attacks in
  continuous-variable quantum cryptography},\ \href@noop {} {\bibfield
  {journal} {\bibinfo  {journal} {Physical review letters}\ }\textbf {\bibinfo
  {volume} {97}},\ \bibinfo {pages} {190502} (\bibinfo {year}
  {2006})}\BibitemShut {NoStop}%
\bibitem [{\citenamefont {Garc{\'\i}a-Patr{\'o}n}\ and\ \citenamefont
  {Cerf}(2006)}]{R61_GAttack2}%
  \BibitemOpen
  \bibfield  {author} {\bibinfo {author} {\bibfnamefont {R.}~\bibnamefont
  {Garc{\'\i}a-Patr{\'o}n}}\ and\ \bibinfo {author} {\bibfnamefont {N.~J.}\
  \bibnamefont {Cerf}},\ }\bibinfo {title} {Unconditional optimality of
  gaussian attacks against continuous-variable quantum key distribution},\
  \href@noop {} {\bibfield  {journal} {\bibinfo  {journal} {Physical review
  letters}\ }\textbf {\bibinfo {volume} {97}},\ \bibinfo {pages} {190503}
  (\bibinfo {year} {2006})}\BibitemShut {NoStop}%
\end{thebibliography}%
\end{document}